\documentstyle[aps,prl,preprint,epsf,floats,epsfig]{revtex}

\newcommand{\beq}{\begin{equation}}
\newcommand{\eeq}{\end{equation}}
\newcommand{\bqa}{\begin{eqnarray}}
\newcommand{\eqa}{\end{eqnarray}}

\def\square{\vcenter{\vbox{\hrule height.4pt
          \hbox{\vrule width.4pt height8pt
          \kern8pt\vrule width.4pt}\hrule height.4pt}}}

\def\sumint{\hbox{$\sum$}\!\!\!\!\!\!\int}

\begin{document}
\preprint{
\vbox{\halign{&##\hfil\cr
        & hep-ph/0107118 \cr
&\today\cr }}}

\title{Solution to the 3-Loop $\Phi$-Derivable Approximation\\
        for Massless Scalar Thermodynamics}

\author{Eric Braaten and Emmanuel Petitgirard}
\address{Physics Department, Ohio State University, Columbus OH 43210,
USA}

\maketitle

\begin{abstract}
We develop a systematic method for solving the 3-loop $\Phi$-derivable 
approximation to the thermodynamics of the massless $\phi^4$ field theory. 
The method involves expanding sum-integrals in powers of $g^2$ and $m/T$, 
where $g$ is the coupling constant, $m$ is a variational mass parameter, 
and $T$ is the temperature. The problem is reduced to one with the single 
variational parameter $m$ by solving the variational equations 
order-by-order in $g^2$ and $m/T$. At the variational point, there are 
ultraviolet divergences of order $g^6$ that cannot be removed by 
any renormalization of the coupling constant. We define a finite 
thermodynamic potential by truncating at $5^{\rm th}$ order in $g$ and 
$m/T$. The associated thermodynamic functions seem to be perturbatively 
stable and insensitive to variations in the renormalization scale.
\end{abstract}

\newpage

\section{Introduction}

The thermodynamic functions for massless relativistic field theories 
at high temperature $T$ can be calculated as weak-coupling expansions
in the coupling constant $g$. They have been calculated explicitly 
through order $g^5$ for the massless $\phi^4$ 
field theory \cite{Parwani-Singh,Braaten-Nieto:scalar}, for QED 
\cite{Coriano-Parwani,Parwani,Andersen}, and for nonabelian 
gauge theories \cite{Arnold-Zhai,Kastening-Zhai,Braaten-Nieto:QCD}. Unless 
the coupling constant is tiny, the weak-coupling expansions are poorly 
convergent and very sensitive to the renormalization scale.
This makes the weak-coupling expansion essentially useless as a quantitative 
tool: the expansion seems to be reliable only when the coupling constant
is so small that the corrections to ideal gas behavior are 
negligibly small.
The physical origin of the instability seems to be
effects associated with screening and quasiparticles. 

A possible solution to this instability problem is to reorganize 
the weak-coupling expansion within a variational framework.  
A variational approximation can be defined by  
a {\it thermodynamic potential} $\Omega$ that depends on a set of 
variational parameters $m_i$.
The free energy and other thermodynamic functions 
are given by the values of $\Omega$ and its derivatives 
at the variational point where $\partial \Omega/\partial m_i = 0$.
A variational approximation is systematically improvable 
if there is a sequence of successive approximations to $\Omega$ 
that reproduce the weak-coupling expansions of the thermodynamic 
functions to successively higher orders in $g$.
One example of a systematically improvable variational approximation
is {\it screened perturbation theory}, which involves a single 
variational mass parameter \cite{K-P-P}.

A variational approach can be useful only if the correct physics 
can be captured by appropriate choices of the variational parameters.
Information about screening and quasiparticle effects is contained
within the exact propagator of the field theory.
The possibility that these effects are responsible for the instability 
of the weak-coupling expansion suggests the use of the 
propagator as a variational function. 
Such a variational formulation was constructed 
for nonrelativistic fermions by Luttinger and 
Ward\cite{Luttinger-Ward} and by Baym \cite{Baym} around 1960. It was 
generalized to relativistic field theories by Cornwall, Jackiw and Tomboulis 
\cite{Cornwall}. In the case of a relativistic scalar field theory, the 
propagator has the form $[P^2 + \Pi(P)]^{-1}$, where $\Pi(P)$ is the 
self-energy which depends on the momentum $P$. The thermodynamic potential 
has the form
\begin{equation}
\Omega_0[\Pi] =
{1 \over 2} 
\sumint_P \left[ \log \left( P^2 + \Pi \right)
	- {\Pi \over P^2 + \Pi} \right]
+ \Phi [\Pi] \,,
\label{thpot}
\end{equation}
where the interaction functional $\Phi[\Pi]$ can be expressed as a sum of 
2-particle-irreducible diagrams.
It is constructed so that the solution to the variational 
equation $\delta \Omega_0/\delta \Pi = 0$ is the exact self-energy,
and the value of $\Omega_0$ at the variational point 
is the exact free energy.
We can obtain a systematically improvable variational approximation
by truncating $\Phi$ at $n$'th order in the loop expansion, 
where $n = 2,3,...$. We refer to this approximation 
as the $n$-loop {\it $\Phi$-derivable approximation}.

One remarkable feature of the 2-loop $\Phi$-derivable approximation is that 
the expression for the entropy reduces at the variational point to the 1-loop 
expression. This was shown for QED by Vanderheyden 
and Baym \cite{Van-Baym}. It was generalized to QCD by Blaizot, Iancu, and 
Rebhan \cite{B-I-R}, who have used it as the basis for a quasiparticle 
model for the thermodynamics of the quark-gluon plasma. An alternative 
quasiparticle model motivated by the 2-loop $\Phi$-derivable approximation
for QCD has been developed by Peshier \cite{Peshier}. One disadvantage of 
these models is that they are based on special properties of the 2-loop 
$\Phi$-derivable approximation. Thus unlike the $\Phi$-derivable 
approximation itself, they are not systematically improvable. 

$\Phi$-derivable approximations are {\it conserving} 
approximations, which means that they are consistent with the conservation 
laws that follow from Noether's theorem. Baym showed that an approximation 
is conserving if and only if it is $\Phi$-derivable for some functional 
$\Phi$ \cite{Baym}. The fact that they are conserving may make 
$\Phi$-derivable approximations particularly useful for nonequilibrium 
problems \cite{Knoll}. They have already proved to be useful for 
nonequilibrium problems in $1+1$ dimensions \cite{Berges}.  

While the $\Phi$-derivable approximation is easily formulated,
it is not so easy to solve. If the solution for the self-energy $\Pi(P)$ is 
independent of the momentum $P$, the $\Phi$-derivable approximation can be 
reduced to a single-parameter variational problem that is easily solved 
numerically. An example is the 2-loop $\Phi$-derivable approximation for the 
massless $\phi^4$ field theory \cite{PKPS}. If the solution for $\Pi(P)$ 
depends on $P$, the variational equation is a nontrivial integral equation. 
What makes it complicated to solve is that it is really a coupled set of 
integral equations for the function $\Pi(2\pi nT, |{\bf p}|)$ corresponding to 
all the Matsubara frequencies $2\pi nT, n=0,\pm 1,\pm 2,...$. An even more 
severe obstacle is  that the thermodynamic potential has ultraviolet 
divergences that vanish at the variational point only if $\Phi$ is 
calculated to all orders. They do not vanish away from the variational 
point, and they do not vanish at the variational point if the loop expansion 
for $\Phi$ is truncated.

In this paper, we solve the 3-loop $\Phi$-derivable
approximation for a massless scalar field theory with a $\phi^4$ interaction. 
In section II, we define the coupling constant for the massless $\phi^4$ 
field theory and give the weak-coupling expansions for several thermodynamic 
quantities. In section III, we summarize the basic features of the 
$\Phi$-derivable approach. In section IV, we solve the 2-loop 
$\Phi$-derivable approximation. We show that if the thermodynamic 
potential is expressed in terms of the true coupling constant, it has 
ultraviolet divergences of order $g^4$. In section V, we solve the 3-loop 
$\Phi$-derivable approximation by expanding sum-integrals systematically 
in powers of $g$ and $m/T$. We show that the thermodynamic potential has 
ultraviolet divergences of order $g^6$. We construct a finite thermodynamic 
potential by adding a term proportional to the square of the variational 
equation and then truncating after terms of $5^{\rm th}$ order in $g$ and 
$m/T$. It defines a stable approximation to the thermodynamic functions that 
is rather insensitive to the choice of renormalization scale. In the 
concluding section, we discuss the outlook for extending our methods to 
gauge theories.

\section{Defining the Theory}

We would like to solve the $\Phi$-derivable approximation to the 
thermodynamics of a
massless scalar field theory with a $\phi^4$ interaction.
The theory can be defined by choosing a regularization scheme 
and specifying the value of the bare coupling constant $g_0$.
The lagrangian for the theory is 
\bqa
{\cal L} \;=\; {1\over2} \partial_{\mu}\phi\partial^{\mu}\phi
- {1\over 24} g_0^2 \phi^4 \;.
\label{barel}
\eqa
We choose to use dimensional regularization in $4-2 \epsilon$
space-time dimensions.  The bare coupling constant $g_0$ then has 
dimensions (mass)$^\epsilon$. We define the renormalized coupling constant 
$g(\mu)$ by the modified minimal subtraction ($\overline{\rm MS}$) 
prescription with renormalization scale $\mu$. For convenience, we also 
define $\alpha = g^2/(4 \pi)^2$ and $\alpha_0 = g_0^2/(4 \pi)^2$.
The relation between the bare and renormalized coupling constants is 
\bqa
\alpha_0 \mu^{-2 \epsilon} &=&
\alpha  + {3 \over 2 \epsilon} \alpha^2
+ \left( {9 \over 4 \epsilon^2} -  {17 \over 12 \epsilon} \right) 
\alpha^3 
+ \left[ {27 \over 8 \epsilon^3} - {119 \over 24 \epsilon^2} 
	+ \left( {145 \over 48} + 2 \zeta(3) \right) {1 \over \epsilon}
	\right] \alpha^4 + ...  \;,
\label{bare-ren}
\eqa
where $\zeta(z)$ is the Riemann zeta function.
In the $\overline{\rm MS}$ scheme, there is usually also a factor of 
$(4 \pi e^{-\gamma})^{2\epsilon}$ on the left side.
We prefer to absorb this factor into the measure of the dimensionally 
regularized integrals.
The beta function for the running coupling constant $\alpha(\mu)$
has been calculated to five-loop order~\cite{beta}.
The first few terms can be obtained by differentiating (\ref{bare-ren})
with respect to $\mu$ and using the fact that $\alpha_0$ 
is independent of $\mu$:
\bqa
\mu{d \ \over d \mu} \alpha \;=\;
-2\epsilon \alpha +\beta(\alpha)\;,
\eqa
where the beta function is
\bqa
\beta(\alpha) \;=\; 3 \alpha^2 - {17\over3} \alpha^3 
+ \left( {145 \over 8} + 12 \zeta(3) \right) \alpha^4 
+ {\cal O}(\alpha^5) \;.
\label{beta}
\eqa

The thermodynamic functions for this theory are known to order $g^5$
\cite{Arnold-Zhai,Parwani-Singh,Braaten-Nieto:scalar}.
The weak-coupling expansion for the free energy density is
\bqa
{\cal F} &=& {\cal F}_{\rm ideal} 
\left[ 1 - {15\over2} {\alpha \over 6} 
+ 60 \left( {\alpha \over 6} \right)^{3/2} 
+ 135 \left( L  -{59\over 45}+{\gamma\over 3}+{4\over 3}{\zeta'(-1)\over 
\zeta(-1)}-{2\over 3}{\zeta'(-3)\over \zeta(-3)} \right) 
\left( {\alpha \over 6} \right)^2 \right.
\nonumber
\\ 
&& \left. \hspace{1cm}
+ 540 \left( 2 \log{\alpha\over6} - 3 L -{5\over 2}+4\log 2-\gamma
+2{\zeta'(-1)\over \zeta(-1)} \right) 
	\left( {\alpha \over 6} \right)^{5/2}
+ {\cal O}(\alpha^3 \log \alpha) \right] \;,
\label{F-weak}
\eqa
where $\alpha=\alpha(\mu)$, $L = \log(\mu/4\pi T)$, $\gamma$ is 
Euler's gamma constant, and ${\cal F}_{\rm ideal}$ is the pressure of the 
ideal gas of a free massless boson:
\begin{equation}
{\cal F}_{\rm ideal} = -{\pi^2\over 90}T^4\,.
\end{equation}
The weak-coupling expansions for the pressure, entropy density, and energy 
density are given by:
\bqa
{\cal P} &=& - {\cal F} \,,
\label{pressure}
\\ 
{\cal S} &=& - {\partial \ \over \partial T} {\cal F}\,,
\label{entropdens}
\\ 
{\cal E} &=& - T^2{\partial \ \over \partial T} \left( {1 \over T} {\cal F} 
\right)\,,
\label{enerdens}
\eqa
where the partial derivatives in (\ref{entropdens}) and (\ref{enerdens}) 
are evaluated with $\alpha(\mu)$ and $\mu$ held fixed. The free energy 
density and other thermodynamic functions satisfy simple renormalization 
group equations:
\begin{equation}
\left[\mu{\partial\over \partial\mu}+\beta(\alpha){\partial \over \partial 
\alpha}\right]{\cal F} = 0 \;.
\label{reneq:free}
\end{equation}  
If the power series expansion for $\beta(\alpha)$ and 
the weak-coupling expansion for ${\cal F}$ are truncated at some order in 
$\alpha$, then the renormalization group equation (\ref{reneq:free}) is 
satisfied only up to higher orders in $\alpha$. 

The renormalization group can be used to improve the weak-coupling expansions 
of the thermodynamic functions. This can be accomplished for the free energy 
(\ref{F-weak}) simply by setting $\mu=c(2\pi T)$, where $c$ is a constant, 
in the logarithm $L$ and in $\alpha(\mu)$. In this case, there are alternative 
thermodynamic definitions of the entropy and energy density in which the 
partial derivatives in (\ref{entropdens}) and (\ref{enerdens}) are replaced 
by total derivatives with respect to $T$. For example, the thermodynamic 
entropy is defined by
\begin{equation}
{\cal S}_{\rm th} = -{d\over dT}{\cal F}\,.
\end{equation}   
It differs from the entropy (\ref{entropdens}) by a term proportional to 
the left side of (\ref{reneq:free}):
\begin{equation}
{\cal S}_{\rm th}-{\cal S} = -{1\over \mu}{d\mu\over dT}\;
\left[\mu{\partial\over \partial\mu}+\beta(\alpha){\partial \over \partial 
\alpha}\right]{\cal F}\,.
\label{diff:entrop}
\end{equation}
The right side is nonzero if the expansions of $\beta$ and ${\cal F}$ in 
powers of $\alpha$ are truncated.

\begin{figure}[t]
\vspace*{-1cm}
\begin{center}
\epsfig{file=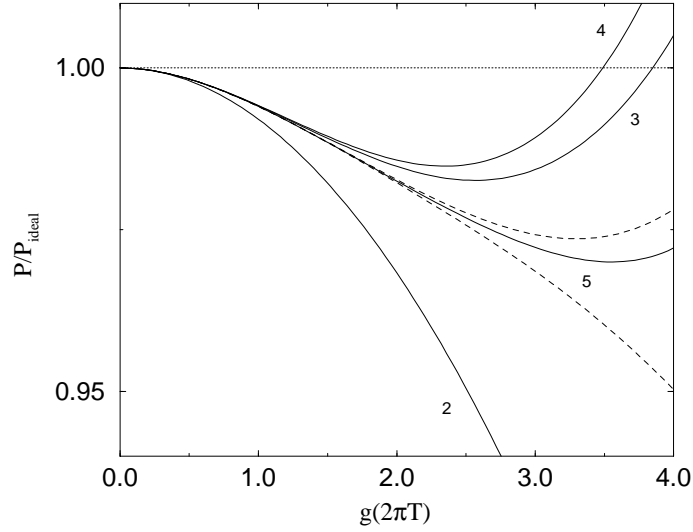,width=8cm,angle=-90}
\end{center}
\vspace*{0.5cm}
\caption{Weak-coupling expansion for the pressure divided by that of the 
ideal gas as a function of $g(2\pi T)$. Solid lines correspond to the 
truncation at order $g^n$, $n=2,3,4,5$ for $\mu =2\pi T$. Dashed lines 
correspond to the truncation at order $g^5$ for $\mu=\pi T$ and $\mu=4\pi T$.}
\end{figure}
The poor convergence of the weak-coupling expansion is illustrated in 
Figure~1, which shows the free energy divided by that of the 
ideal gas as a function of $g(2\pi T)$. The solid curves are the 
weak-coupling expansion with renormalization scale $\mu_T=2\pi T$ truncated 
after orders $g^n$, $n=2,3,4,5$. They show no sign of converging except at 
extremely small values of $g$. The sensitivity of the weak-coupling expansion 
to the renormalization scale is also illustrated in Figure~1. The dashed lines 
are the weak-coupling expansion through order $g^5(\mu)$ for $\mu=\pi T$, 
$\mu=4\pi T$ plotted as a function of $g(\mu_T)$. The relation between 
$g(\mu)$ and $g(\mu_T)$ is obtained by integrating numerically from $2\pi T$ 
to $\mu$ using the beta function (\ref{beta}) truncated after the $\alpha^4$ 
term. Note that $g=4$, which corresponds to $\alpha=0.1013$, represents 
relatively strong coupling in the sense that it is close to the Landau pole. 
If $g(2\pi T)=4$, the Landau pole at which the running coupling constant 
defined by the 3-loop beta function diverges is $\mu=12.15(2\pi T)$. 

Another observable that has been calculated to relatively high order 
is the screening mass \cite{Braaten-Nieto:scalar}.
It is known up to corrections of order $g^5$:
\bqa
m_s^2 &=& {\alpha \over 6} (2 \pi T)^2
\left[ 1 - 6 \left({\alpha \over 6}\right)^{1/2} 
+ 6 \left[ 2 \log {\alpha \over 6} - 3 L -1+8\log 2 -\gamma+2
{\zeta'(-1)\over \zeta(-1)} \right] 
	{\alpha \over 6}\right.\nonumber
\\
&&\hspace{2cm}\left. 
+ {\cal O}(\alpha^{3/2}\log \alpha ) \right]\;.
\label{ms-weak}
\eqa
%
\begin{figure}[t]
\vspace*{-1cm}
\begin{center}
\epsfig{file=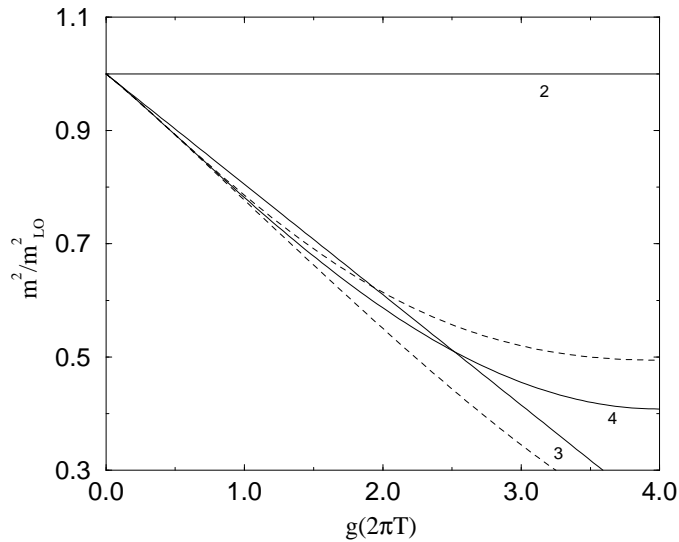,width=8cm,angle=-90}
\end{center}
\vspace*{0.5cm}
\caption{Weak-coupling expansion for the screening mass divided by the 
leading order result $m^2_{\rm LO}=g^2(2\pi T)T^2/24$ as a function of 
$g(2\pi T)$. Solid lines correspond to the truncation at order $g^n$, 
$n=2,3,4$ for $\mu =2\pi T$. Dashed lines correspond to the truncation 
at order $g^4$ for $\mu=\pi T$ and $\mu=4\pi T$.}
\end{figure}
The convergence of the weak-coupling expansion is illustrated in 
Figure~2, which shows $m_s^2$ divided by the leading order expression 
$m_{\rm LO}=g^2(\mu_T)T^2/24$ as a function of $g(2\pi T)$. The solid lines 
are the weak-coupling expansion (\ref{ms-weak}) truncated after the term of 
order $g^n$, $n=2,3,4$. The weak-coupling expansions to orders $g^3$ and 
$g^4$ agree only at extremely small values of $g$. The expansion to order 
$g^4$ is also very sensitive to the renormalization scale as illustrated in 
Figure~2. The dashed lines show $m_s^2/m_{\rm LO}^2$, with $m_s^2$ truncated 
after the $g^4(\mu)$ term as a function of $g(2\pi T)$ for $\mu=\pi T$ and 
$\mu=4\pi T$.

\section{$\Phi$-derivable approximations}

$\Phi$-derivable approximations are based on the skeleton expansion 
of the free energy \cite{Luttinger-Ward}, which expresses it in terms of 
exact propagators. The exact propagator $D(P)$ for a massless scalar 
theory has the form
\begin{equation}
D(P) = {1 \over P^2 + \Pi (P)} \,.
\end{equation}
The skeleton expansion for the free energy defines a functional $\Omega [D]$ 
of the propagator that we will refer to as the {\it thermodynamic potential}. 
The exact propagator satisfies the variational equation 
$\delta \Omega/\delta D = 0$. The free energy is therefore the variational 
minimum of the functional $\Omega [D]$:
\begin{equation}
{\cal F} = \Omega [D] \bigg|_{\delta \Omega/\delta D = 0} \,.
\label{vary:1}
\end{equation}
The thermodynamic potential can be expressed in the form
\begin{equation}
\Omega [D] =
{1 \over 2} \sumint_P \log \left(P^2 + \Pi (P)\right)
- {1 \over 2} \sumint_P \Pi (P) {1 \over P^2 + \Pi (P)}
+ \Phi [D] \,,
\label{thpot:phi}
\end{equation}
where the functional $\Phi[D]$ contains all the effects of interactions. 
It can be expressed as a sum of 2-particle-irreducible Feynman diagrams. 
The variational equation $\delta \Omega /\delta D = 0$ can be written
\begin{equation}
\Pi (P) = 2 {\delta \Phi [D] \over \delta D (P)} \,.
\label{vary:2}
\end{equation}
The variational derivative is calculated with the bare coupling constant 
$\alpha_0$ and the ultraviolet cutoff parameter $\epsilon$ held fixed or, 
equivalently, with the renormalized coupling constant $\alpha$ and the 
renormalization scale $\mu$ held fixed.  

An approximation is called $\Phi$-derivable if it can be obtained by
making the same approximation for the functional $\Phi$ in both
(\ref{thpot:phi}) and (\ref{vary:2}) \cite{Baym}.  
$\Phi$-derivable approximations
have several attractive features.  One such feature is {\it
self-consistency}:  the thermodynamic potential (\ref{thpot:phi}), which
includes the effects of interactions through the $\Phi$ term, is
evaluated with a propagator that takes into account those same
interactions through the variational equation (\ref{vary:2}).  Another
attractive feature is {\it thermodynamic consistency}. The entropy density 
can be expressed as
\begin{equation}
{\cal S} =
\left( - {\partial \ \over \partial T} \Omega [D] \right )
\Bigg|_{\delta \Omega/\delta D = 0}\;,
\label{entropy:1}
\end{equation}
where $\partial / \partial T$ refers to the partial derivative with 
$D(P)$, $\alpha_0$, and $\epsilon$ held fixed. The thermodynamic entropy 
can be written
\begin{equation}
{\cal S_{\rm th}} = - {d \ \over dT}
\left( \Omega [D] \bigg|_{\delta\Omega/\delta D = 0} \right) \,,
\label{entropy:2}
\end{equation}
where the derivative is evaluated with $\alpha_0$ and $\epsilon$ fixed. 
The equivalence of (\ref{entropy:1}) and (\ref{entropy:2}) follows from 
the variational equation $\delta \Omega/\delta D = 0$.  

The $n$-loop $\Phi$-derivable approximations defined by truncating $\Phi$ 
at successively higher orders in the bare coupling constant, or equivalently 
the loop expansion, are particularly useful. They define a systematically 
improvable variational approximation. If $\Phi$ is truncated at order 
$g_0^{2(n-1)}$, the free energy will after renormalization include correctly 
all terms in the weak-coupling expansion through order $g^{2n-1}$, as well 
as an infinite series of higher terms in $g$.  Thus each successive
approximation will have an error that is suppressed parametrically
by one higher order in $g^2$.  The propagator for each approximation
is given by the self-consistent solution to a variational equation
that takes into account the screening and quasiparticle effects 
characteristic of a plasma. Thus if the poor convergence of the 
weak-coupling expansion is due to treating plasma effects as a 
perturbation, the $n$-loop $\Phi$-derivable approximations could provide the 
solution to this problem. 

Unfortunately, $\Phi$-derivable approximations have a problem that has limited 
their usefulness in relativistic field theories. The problem is that the 
thermodynamic potential $\Omega$ is generally ultraviolet divergent. There 
are two classes of ultraviolet divergences. One class consists of ultraviolet 
divergences that vanish at the variational point. They provide a serious 
complication for devising numerical methods for minimizing the 
thermodynamical potential. The other class of ultraviolet divergences are 
those that survive at the variational point. They arise from having truncated 
$\Phi$ at $n^{\rm th}$ order in the loop expansion. Renormalization theory 
guarantees that the weak-coupling expansion for the free energy will be 
finite through order $g^{2n-1}$, but there will generally be ultraviolet 
divergences at order $g^{2n}$. In the 2-loop $\Phi$-derivable approximation 
for the massless $\phi^4$ field theory, these ultraviolet divergences can be 
hidden by introducing a coupling constant that has the wrong beta function. 
However in the 3-loop $\Phi$-derivable approximation, these ultraviolet 
divergences must be dealt with.

\section{Two-Loop $\Phi$-derivable approximation}
\label{Two-loop}

In the 2-loop $\Phi$-derivable approximation, the solution for the exact 
self-energy is independent of the momentum. We anticipate this fact by
denoting the exact self-energy by $m^2$. The variable $m$ can be 
interpreted as the screening mass. The solution to the 2-loop 
$\Phi$-derivable approximation has been given previously in 
Refs.~\cite{PKPS} and \cite{B-I-R}. We first solve the 2-loop 
$\Phi$-derivable approximation exactly in terms of a coupling 
constant $\bar{g}(\mu)$ that has the wrong beta function. We show that the 
expansion of the thermodynamic potential in powers of $m/T$ converges much 
more rapidly to the exact answer than the weak-coupling expansion in powers 
of $\bar{g}$. We then show that when the solution is expressed in terms of 
the true coupling constant $g$, there are unavoidable ultraviolet divergences 
at order $g^4$.   

The thermodynamic potential in the 2-loop $\Phi$-derivable approximation is
\begin{equation}
\mu^{2 \epsilon} \Omega_0 (m) = {1 \over 2} \sumint_P \log (P^2 + m^2)
- {1 \over 2} m^2 \sumint_P {1 \over P^2 + m^2}
+ {1 \over 8} (g_0 \mu^{-\epsilon})^2
        \left( \sumint_P {1 \over P^2 + m^2} \right)^2 \;,
\label{thpot2:1}
\end{equation}
where $g_0$ is the bare coupling constant.
Our definition of the dimensionally regularized sum-integral 
is given in (\ref{sumint-def}) and includes a factor of $\mu^{2 \epsilon}$ 
in the measure. Each term in (\ref{thpot2:1}) has an overall multiplicative 
factor of $\mu^{2 \epsilon}$, and therefore $\Omega_0(m)$ is independent of 
$\mu$ for fixed $g_0$ and $\epsilon$. This factor of $\mu^{2\epsilon}$ 
is convenient, because it gives every term in (\ref{thpot2:1}) the 
dimensions (mass)$^4$, even for $\epsilon \neq 0$. The variational 
equation $\partial \Omega/\partial m = 0$ reduces to the simple gap 
equation
\begin{equation}
m^2 = {1 \over 2} (g_0\mu^{-\epsilon})^2 \sumint_P {1 \over P^2 + m^2}
\,.
\label{gap2:1}
\end{equation}
The 2-loop $\Phi$-derivable approximation for the free energy density
${\cal F}$ is obtained by inserting the solution to the gap equation
(\ref{gap2:1}) into the thermodynamic potential (\ref{thpot2:1}).

\subsection {Finite thermodynamic potential}

The sum-integrals that appear in the thermodynamic potential (\ref{thpot2:1}) 
and in the gap equation (\ref{gap2:1}) are ultraviolet divergent. Ideally, we 
would like to have a thermodynamic potential $\Omega(m)$ that is a 
finite function of the variational parameter $m$. To construct 
such a function, we begin by making the ultraviolet divergences explicit as 
poles in $\epsilon$. The sum-integrals can be expressed in the form 
\begin{eqnarray}
\sumint_P\log(P^2+m^2) &=& -{\pi^2\over 45}T^4\hat \mu^{2\epsilon}
\left\{a(\epsilon)-15f(\hat{m}^2,\epsilon) +45
\left[{1\over 2\epsilon}+{3\over 4}b(\epsilon)\right]\hat{m}^{4-2\epsilon}
\right\}\,,
\label{sumlog}
\\
\sumint_P{1\over P^2+m^2} &=& {1\over 12}T^2 \hat \mu^{2\epsilon}
\left\{f'(\hat{m}^2,\epsilon)-3(2-\epsilon)\left(
{1\over 2\epsilon}+{3\over 4}b(\epsilon)\right)\hat{m}^{2-2\epsilon}
\right\}\,,
\label{sumprop}
\end{eqnarray} 
where $\hat{m}=m/(2\pi T)$, $\hat{\mu}=\mu/(2\pi T)$, and $a(\epsilon)$ 
and $b(\epsilon)$ are constants that approach $1$ in the limit 
$\epsilon\to 0$:
\begin{eqnarray}
a(\epsilon) &=& {e^{\gamma\epsilon}(2\pi)^{2\epsilon}\Gamma({5\over 2})
\Gamma(4-2\epsilon)\zeta(4-2\epsilon)\over \Gamma({5\over 2}-\epsilon)
\Gamma(4)\zeta(4)}\,, 
\\
b(\epsilon) &=& {4\over 3}\left[{e^{\gamma\epsilon}\Gamma(1+\epsilon)
\over \epsilon(1-\epsilon)(2-\epsilon)}-{1\over 2\epsilon}\right]\,.
\end{eqnarray}
The function $f(\hat{m}^2,\epsilon)$ is given by the integral
\begin{equation}
f(\hat{m}^2,\epsilon)=-16{e^{\gamma\epsilon}\Gamma({5\over 2})\over
\Gamma({5\over 2}-\epsilon)}\int_0^{\infty}dx\left[{x^{4-2\epsilon}\over 
(x^2+\hat{m}^2)^{1/2}}
{1\over e^{2\pi(x^2+\hat{m}^2)^{1/2}}-1}-{x^{3-2\epsilon}\over e^{2\pi x}-1} 
\right]\,,
\label{f:eps}
\end{equation}
and $f'(\hat{m}^2,\epsilon)$ is its derivative with respect to $\hat{m}^2$. 
In the limit $\hat{m}\to 0$, $f(\hat{m},0)$ approaches $\hat{m}^2$. 
The sum-integral (\ref{sumprop}) can be obtained simply by 
differentiating (\ref{sumlog}) with respect to $m^2$. Inserting 
(\ref{sumlog}) and (\ref{sumprop}) into (\ref{thpot2:1}), the thermodynamic 
potential reduces to 
\begin{eqnarray}
{\mu^{2\epsilon}\Omega_0(m)\over {\cal F}_{\rm ideal}} &=& \left[a(\epsilon)
-15f(\hat{m}^2,\epsilon)+15\hat{m}^2f'(\hat{m}^2,\epsilon)\right]\hat 
\mu^{2\epsilon}-45(1-\epsilon)\left({1\over 2\epsilon}+{3\over 4}b(\epsilon)
\right) \hat \mu^{2\epsilon}\hat{m}^{4-2\epsilon}
\nonumber
\\
&& 
-{5\over 4}\alpha_0\mu^{-2\epsilon}\left[f'(\hat{m}^2,\epsilon)-3
(2-\epsilon)\left({1\over 2\epsilon}+{3\over 4}b(\epsilon)\right)
\hat{m}^{2-2\epsilon}\right]^2\hat \mu^{4\epsilon}\,.
\label{thpot2:f}
\end{eqnarray}
The gap equation (\ref{gap2:1}) reduces to
\begin{equation}
\hat{m}^2={\alpha_0\mu^{-2\epsilon}\over 6}\left[f'(\hat{m}^2,\epsilon)-
3(2-\epsilon)\left({1\over 2\epsilon}+{3\over 4}b(\epsilon)\right)
\hat{m}^{2-2\epsilon}\right]\hat \mu^{2\epsilon}\,.
\label{gap2:f}
\end{equation}

We wish to construct a finite thermodynamic potential $\Omega$ that has 
the same value at the variational point as (\ref{thpot2:f}). The gap equation 
(\ref{gap2:f}) can be expressed in the form $\hat{m}^2-\hat{G}=0$. We can add 
any multiple of $(\hat{m}^2-\hat{G})^2$ to (\ref{thpot2:f}) without changing 
its value at the variational point. We can use this freedom to cancel the 
double pole in $\epsilon$ in (\ref{thpot2:f}). Adding $45(\hat{m}^2-\hat{G})
^2/(\alpha_0\mu^{-2\epsilon})$ to (\ref{thpot2:f}), the resulting function 
reduces to
\begin{equation}
{\mu^{2\epsilon}\Omega(m)\over {\cal F}_{\rm ideal}}=\left[a(\epsilon)-15
f(\hat{m}^2,\epsilon)\right]\hat \mu^{2\epsilon}+{45\over \alpha_0
\mu^{-2\epsilon}}\hat{m}^4+45\left({1\over 2\epsilon}+{3\over 4}b(\epsilon)
\right)\hat \mu^{2\epsilon} \hat{m}^{4-2\epsilon}
\label{thpot2:ff}
\end{equation}
This thermodynamic potential will be finite as $\epsilon\to 0$ provided that 
the ``renormalized coupling constant'' $\bar{\alpha}(\mu)$ defined by
\begin{equation}
{1\over \bar \alpha(\mu)}={1\over \alpha_0\mu^{-2\epsilon}}
+{1\over 2\epsilon}
\label{coup:f}
\end{equation}
is finite as $\epsilon\to 0$. This equation does not define the value of 
$\bar{\alpha}(\mu)$ at any specific scale $\mu$, because (\ref{coup:f}) 
does not have a smooth limit as $\epsilon\to 0$. However its dependence on 
$\mu$ is well defined. Differentiating (\ref{coup:f}) with respect to $\mu$, 
we obtain 
\begin{equation}
\mu{d\over d\mu}\bar \alpha(\mu)=-2\epsilon\bar \alpha(\mu)+
\bar \alpha^2(\mu)\,.
\label{beta:f}
\end{equation}
This equation has a smooth limit as $\epsilon\to 0$, so the $\mu$-dependence 
of $\bar{\alpha}(\mu)$ is well-defined. Note that the coefficient of 
$\bar{\alpha}^2$ in (\ref{beta:f}) differs from the coefficient of 
$\alpha^2$ in the true beta function (\ref{beta}) by a factor of 3.

Using (\ref{coup:f}) to eliminate $\alpha_0$ from (\ref{thpot2:ff}) in 
favor of $\bar{\alpha}(\mu)$, we obtain a finite thermodynamic potential 
in the limit $\epsilon\to 0$:
\begin{equation}
{\Omega(m)\over {\cal F}_{\rm ideal}}= 1-15\left[f(\hat{m}^2)
-3\left({1\over \bar \alpha}+\log{\mu\over m}+{3\over 4}
\right)\hat{m}^4\right]\,,
\label{thpot2:fff}
\end{equation}
where the function $f(\hat{m}^2)$ is now the limit as $\epsilon\to 0$ of 
(\ref{f:eps}):
\begin{equation}
f(\hat{m}^2)=-16\int_0^{\infty}dx\left({x^4\over \sqrt{x^2+\hat{m}^2}}
{1\over e^{2\pi(x^2+\hat{m}^2)^{1/2}}-1}-{x^3\over e^{2\pi x}-1} \right)\,.
\end{equation}
Upon varying (\ref{thpot2:fff}) with respect to $m^2$, we obtain the gap 
equation 
\begin{equation}
\hat{m}^2={\bar \alpha\over 6}\left[f'(\hat{m}^2)-6\left(\log
{\mu\over m}+{1\over 2}\right)\hat{m}^2\right]\,.
\label{gap2:fff}
\end{equation}
The free energy ${\cal F}(T)$ in the 2-loop $\Phi$-derivable approximation is 
obtained by solving (\ref{gap2:fff}) for $\hat{m}$ and inserting the 
solution into (\ref{thpot2:fff}).

The thermodynamic potential (\ref{thpot2:fff}) is independent of the scale 
$\mu$, because it satisfies the renormalization group equation
\begin{equation}
\left[\mu{\partial \over \partial\mu}+\bar \alpha^2{\partial \over 
\partial \bar \alpha}\right]\Omega(m) = 0\,.
\end{equation}
However the $\Phi$-derivable free energy still has an ambiguity associated 
with the renormalization scale. The ambiguity arises because (\ref{coup:f}) 
determines the dependence of $\bar \alpha(\mu)$ on $\mu$ but it does not 
define the value of $\bar \alpha(\mu)$ at any scale $\mu$. The value of 
$\bar{\alpha}(\mu)$ becomes well-defined only after some initial value 
$\bar{\alpha}(\mu_0)$ is specified. A convenient choice for $\mu_0$ is 
the scale at which $\bar{\alpha}$ has the same value as the true coupling 
constant $\alpha$:
\begin{equation}
\bar \alpha(\mu_0) = \alpha(\mu_0) \,.
\label{match}
\end{equation}
Since $\bar{\alpha}$ and $\alpha$ have different beta functions,
their values are different at other values of $\mu$:
\begin{equation}
\bar \alpha(\mu) \;=\; \alpha(\mu) 
- 2 \log{\mu \over \mu_0} \alpha^2(\mu)
+ \left[ 4 \log^2{\mu \over \mu_0} + {17\over3}\log{\mu \over \mu_0} \right]
	\alpha^3(\mu)
+ {\cal O}(\alpha^4) \,.
\label{albare-al}
\end{equation}
The ambiguity in the matching scale $\mu_0$ is the price that must be paid 
for using a coupling constant with the wrong beta function. A reasonable 
choice for the matching scale is $\mu_0=2\pi T$. However in estimating the 
errors of the $\Phi$-derivable approximation, one should take into account 
the ambiguity in $\mu_0$, perhaps by allowing it to vary by a factor of 2 
around the preferred value.

\subsection{Weak-coupling Expansion}

The two-loop free energy in the $\Phi$-derivable approximation can be 
expanded systematically as a weak-coupling expansion in powers of 
$\bar{\alpha}^{1/2}$. One can obtain a sequence of analytic approximation to 
the free energy by truncating the expansion. Unfortunately we will find that 
the series converges rather slowly.

To express the free energy as an expansion in powers of $\bar{\alpha}^{1/2}$, 
we need to expand the solution to the gap equation (\ref{gap2:fff}) in powers 
of $\bar{\alpha}^{1/2}$ and insert it into (\ref{thpot2:fff}). We first need 
to expand the function $f(\hat{m}^2)$ in powers of $\hat{m}$. The expansion, 
which is derived in Appendix~\ref{app:movert}, has the form 
\begin{equation}
f(\hat{m}^2)=\hat{m}^2-4\hat{m}^3-3\left(\log{\hat{m}\over 2}-{3\over 4}
+\gamma\right)\hat{m}^4+\sum_{n=3}^{\infty}f_n\hat{m}
^{2n}\,,
\label{we:1}
\end{equation}
where the coefficients $f_n$ are 
\begin{equation}
f_n=(-1)^{n+1}{6\Gamma(n-{3\over 2})\zeta(2n-3)\over \Gamma({1\over 2})n!}\,.
\label{we:2}
\end{equation}
Note that the only odd power of $\hat{m}$ is the $\hat{m}^3$ term. The 
asymptotic behavior of the coefficients as $n\to \infty$ is $f_n\to (-1)
^{n+1}(6/\sqrt{\pi})n^{-5/2}$. Using the ratio test, we can infer that the 
region of convergence of the power series in (\ref{we:1}) is $|m|<2\pi T$. 

Inserting the expansion (\ref{we:1}) into (\ref{thpot2:fff}), the 
thermodynamic potential becomes
\begin{equation}
{\Omega(m)\over {\cal F}_{\rm ideal}} = 1-15\left\{ \hat{m}^2-4\hat{m}^3-3
\left({1\over \bar \alpha}+L+\gamma\right)\hat{m}^4+\sum_{n=3}^{\infty}f_n
\hat{m}^{2n} \right\} \,,
\label{thpot:fn}
\end{equation} 
where $L=\log(\mu/4\pi T)$. Differentiating the expansion 
(\ref{we:1}) and inserting it into (\ref{gap2:fff}), the gap equation becomes
\begin{equation}
\hat{m}^2={\bar \alpha \over 6}\left\{1-6\hat{m}-6(L+\gamma)\hat{m}^2
+\sum_{n=3}^{\infty}nf_n\hat{m}^{2n-2}\right\}\,.
\label{gap:fn}
\end{equation}
Using the ratio test, we can infer that the regions of convergence
of the series in (\ref{thpot:fn}) and (\ref{gap:fn}) is $|m| < 2 \pi T$. 
As long as $m$ is inside this radius of convergence, the functions of $m/T$ 
on the right can be evaluated simply by adding terms in the series until 
the desired numerical accuracy is achieved. The first few terms in the 
series for (\ref{gap:fn}) are
\begin{eqnarray}
\hat{m}^2 &=& {\bar \alpha \over 6}
\left \{ 1 - 6 \hat{m}
- 6 (L+\gamma) \hat{m}^2
+ 1.803 \hat{m}^4
- 0.778 \hat{m}^6  + \ldots
\right\}.
\label{gap2:5}
\end{eqnarray}
For asymptotically weak coupling, the solution is $m = \bar{g}T / \sqrt {24}$. 
The radius of convergence in $m$ translates into $|\bar{g}| < 4 \pi \sqrt 6 
\simeq 30.8$.

The solution to the gap equation (\ref{gap:fn}) up to errors of order 
$\bar{\alpha}^{5/2}$ is 
\begin{eqnarray}
\hat{m}^2 &=& {\bar \alpha \over 6}
\left\{ 1 - 6\left({\bar \alpha \over 6}\right)^{1\over 2} - 6(L+\gamma -3) 
{\bar \alpha \over 6}
+ 27(2L+2\gamma -1) \left({\bar \alpha \over 6}\right)^{3/2}
\right.
\nonumber
\\
&& \hspace{1cm} \left .
+ 36\left( L^2 - (6-2\gamma)L-3.080\right) \left({\bar \alpha 
\over 6}\right)^2 +{\cal O} (\bar \alpha^{5/2}) \right \} \,,
\label{m2:1}
\end{eqnarray}
where $\bar \alpha=\bar \alpha(\mu)$ and $L=\log(\mu/4\pi T)$. 
The weak-coupling expansion for the two-loop $\Phi$-derivable free energy is 
obtained by inserting the expansion (\ref{m2:1}) into (\ref{thpot:fn}) 
and expanding in powers of $\bar{\alpha}^{1/2}$. The expansion up to errors 
of order $\bar{\alpha}^{7/2}$ is 
\begin{eqnarray}
{1\over 15}\left( {{\cal F}\over {\cal F}_{\rm ideal}}-1\right) & = & 
- {1 \over 2} {\bar \alpha \over 6}+ 4 \left({\bar \alpha \over 6}\right)^{3/2}
+ 3 (L+\gamma -6) \left({\bar \alpha \over 6}\right)^2
- 18 (2L+2\gamma -3) \left({\bar \alpha \over 6}\right)^{5/2}
\nonumber
\\
&& 
- 18 \left( L^2 - (12-2\gamma) L -0.560 \right) \left( {\bar 
\alpha \over 6}\right)^3 + {\cal O}(\bar \alpha^{7/2}) \;.
\label{free2:4}
\end{eqnarray}
%

\begin{figure}[t]
\vspace*{-1cm}
\begin{tabular}{cc}
\epsfig{file=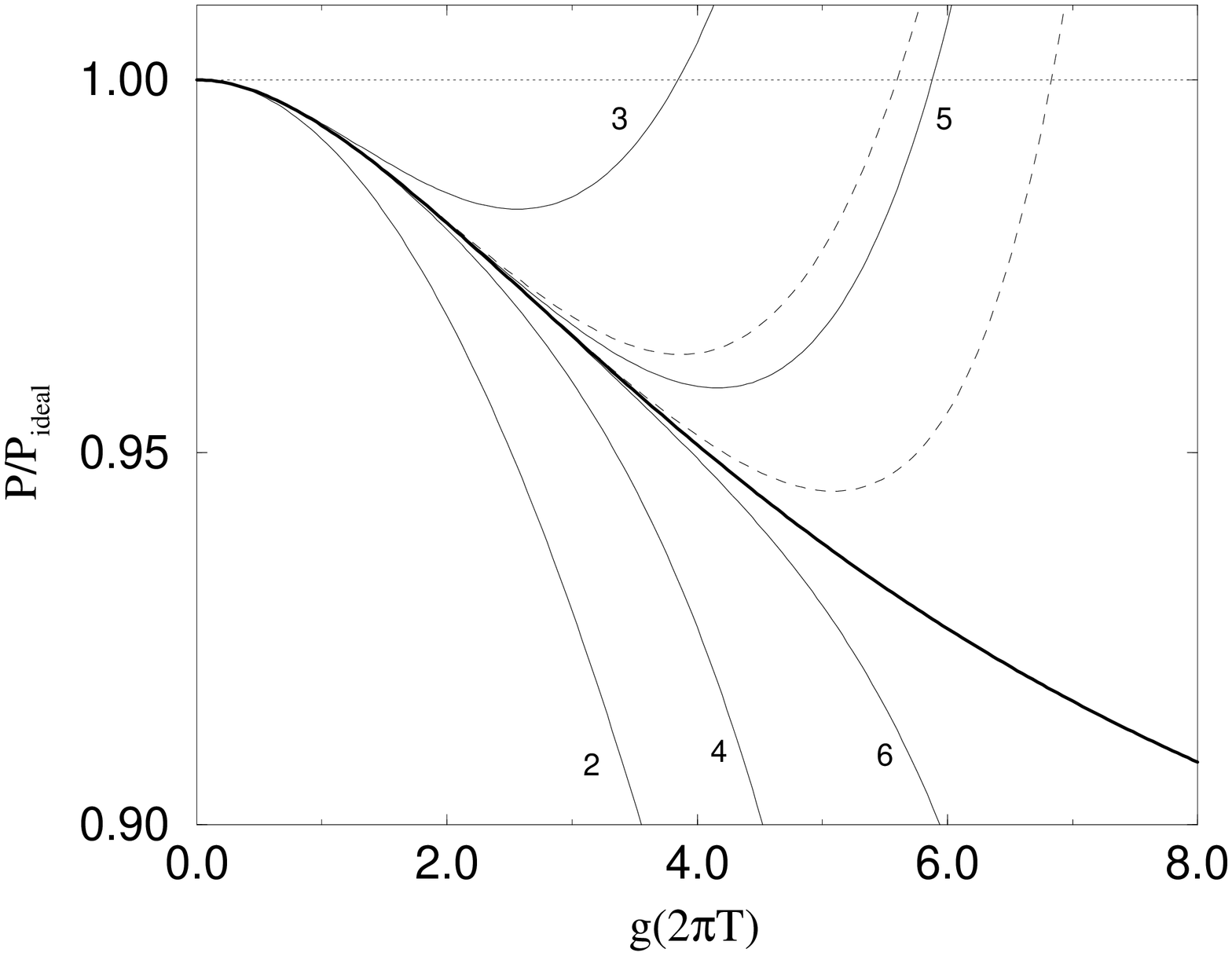,width=7cm,angle=-90}
&
\epsfig{file=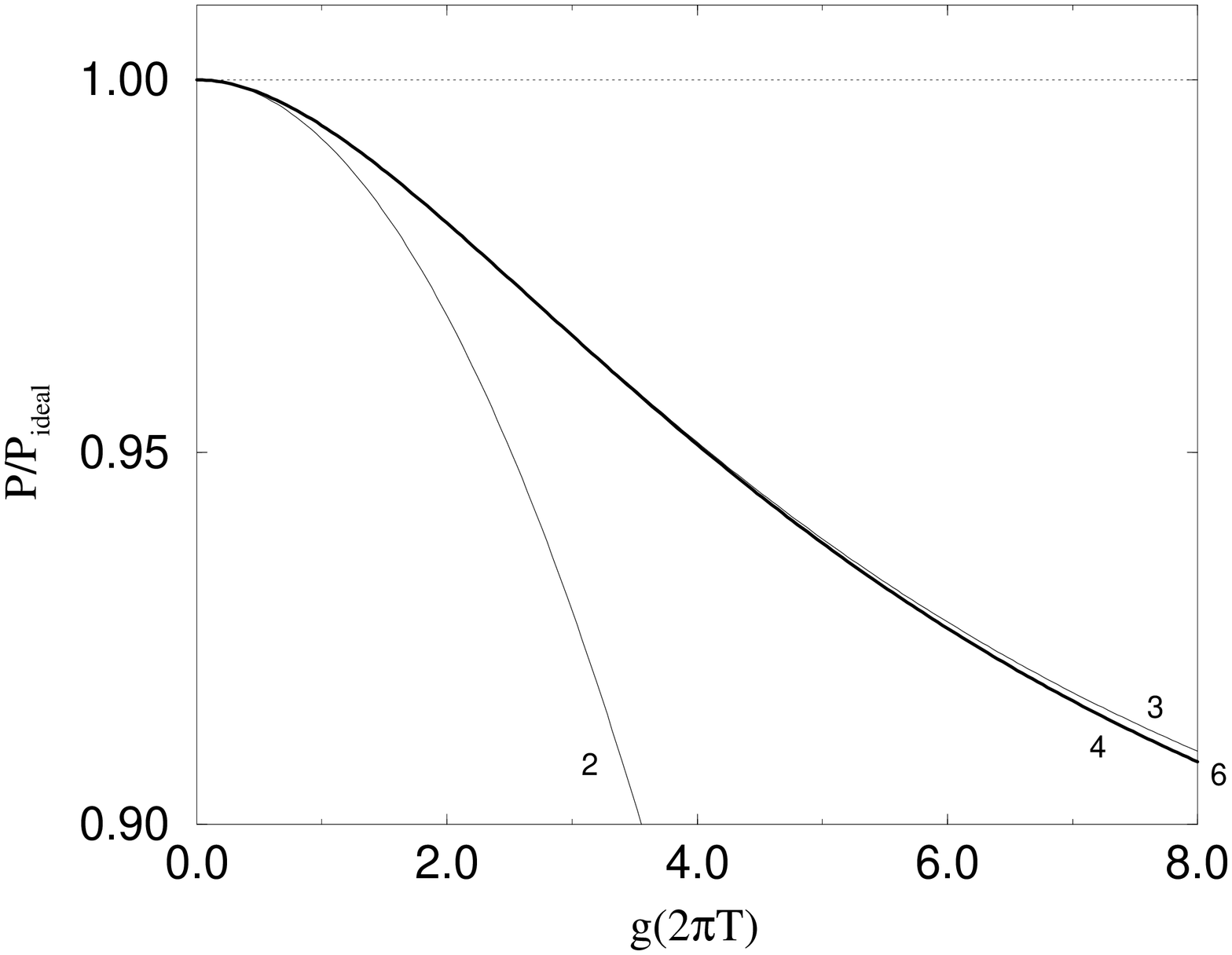,width=7cm,angle=-90}
\vspace*{0.5cm}
\\
\vspace*{0.5cm}
(a) & (b)
\end{tabular}
\caption{2-loop $\Phi$-derivable approximation to the pressure divided by 
that of the ideal gas (heavy solid line) as a function of $\bar{g}(2\pi T)$. 
In Figure~(a), this is compared to the weak-coupling expansion truncated at 
orders $g^n$, $n=2,3,4,5,6$. In Figure~(b), it is compared to the $m/T$ 
expansion truncated at orders $g^n$, $n=2,3,4,6$.}
\end{figure}
The convergence behavior of the weak-coupling expansion for the 2-loop 
$\Phi$-derivable approximation is illustrated in Figure~3, which shows the 
free energy divided by that of the ideal gas as a function of 
$\bar{g}(2\pi T)$. The exact result is shown as a 
heavy solid line. The light solid lines show the weak-coupling expansions for 
$\mu=2\pi T$ truncated after the terms of order $\bar{g}^n$, $n=2,3,4,5$. 
The weak-coupling expansion converges to the exact answer, which is 
reflected in the fact that successive approximations track the exact answer 
out to higher values of $\bar{g}$. However the convergence is relatively 
slow. While the exact answer is completely independent of the renormalization 
scale $\mu$, the weak-coupling expansion has a strong sensitivity to $\mu$ as 
illustrated in Figure~3a. The dashed lines are the weak-coupling expansions 
for $\mu=\pi T$ and $\mu=4\pi T$ truncated after the $\bar{g}^5(\mu)$ term 
as a function of $\bar{g}(2\pi T)$. The relation between $\bar{g}(\mu)$ and 
$\bar{g}(2\pi T)$ is given by the exact solution to the renormalization 
group equation:
\begin{equation}
\bar\alpha(\mu)={\bar\alpha(2\pi T)\over 1-\bar\alpha(2\pi T)
\log(\mu/2\pi T)}\,.
\end{equation}   
Note that the Landau pole appears at $\mu=e^{1/\bar{\alpha}(2\pi T)}(2\pi T)$. 
For $\bar{g}(2\pi T)=8$, it appears at $\mu=11.79(2\pi T)$. Thus 
$\bar{g}=8$ in the 2-loop $\Phi$-derivable approximation is roughly as 
close to the Landau pole as $g=4$, where $g$ is the running coupling 
constant defined by the 3-loop beta function of the $\phi^4$ theory.

\subsection{Expansion in $m/T$}

The slow convergence of the weak-coupling expansion provides motivation 
for developing a better sequence of successive approximations to the 
$\Phi$-derivable free energy. We would like an approximation scheme 
that is applicable not only to the two-loop $\Phi$-derivable approximation, 
but can be extended to higher loops. We will develop an approximation scheme 
that is based on expanding sum-integrals in powers of $m/T$. 

The expansion of the finite thermodynamic potential $\Omega(m)$ in powers of 
$\hat{m}=m/(2\pi T)$ is given in (\ref{thpot:fn}). We can define a series of 
successive approximations $\Omega^{(n)}(m)$ by truncating after terms of 
$n$-th order in $\bar{g}$, with $m$ treated as order $\bar{g}$. For example, 
the thermodynamic potential truncated after the term of order $\bar{g}^6$ is
\begin{equation}
{1\over 15}\left( {\Omega^{(6)}(m)\over {\cal F}_{\rm ideal}}-1\right) = 
{3\hat m^4\over \bar \alpha}-\hat m^2+4\hat m^3+3\left( L+\gamma \right)
\hat m^4-{1\over 2}\zeta(3)\hat m^6 \,.
\label{free:n=6}
\end{equation}
The corresponding gap equation is 
\begin{equation}
\hat{m}^2 = {\bar \alpha \over 6}\left[1-6\hat{m}-6(L+\gamma)\hat{m}^2
+{3\over 2}\zeta(3)\hat{m}^4 \right]\,.
\label{gap:N6}
\end{equation}

The first few of these successive approximations to the free energy can be 
calculated analytically. The $n=2$ truncation includes only the first two 
terms of the right side of (\ref{free:n=6}). The solution to the gap equation 
is simply $\hat{m}=\sqrt{\bar{\alpha}/6}$. The resulting expression for the 
free energy is 
\begin{equation}
{{\cal F}^{(2)}\over {\cal F}_{\rm ideal}}=1-{5\over 4}\bar{\alpha}\,.
\end{equation}
The $n=3$ truncation includes also the $\hat{m}^3$ term in (\ref{free:n=6}). 
The gap equation is a quadratic equation whose solution is 
\begin{equation}
\hat{m} \;=\; 
 \sqrt{ {1\over 6} \bar \alpha + {1 \over 4} \bar \alpha^2 }
	-{1 \over 2} \bar \alpha .
\label{quadra-sol}
\end{equation}
The resulting expression for the free energy  is
\begin{equation}
{{\cal F}^{(3)}\over {\cal F}_{\rm ideal}} = 1-{5\over 4}\bar \alpha
\left[ 1+6\bar \alpha+6\bar \alpha^2-4\left(2+3\bar \alpha\right)
\sqrt{{\bar \alpha\over 6}+{\bar \alpha^2\over4}} \right] \,.
\label{free:sqrt} 
\end{equation} 
In the strong coupling limit $\bar{\alpha}\to \infty$, the solutions for the 
$n=3$ truncation have finite limits: $\hat{m}\to {1\over 6}$ and ${\cal F}
^{(3)}/{\cal F}_{\rm ideal}\to {31\over 36}$. Note that solutions 
(\ref{quadra-sol}) and (\ref{free:sqrt}) both depend on the renormalization 
scale $\mu$ through $\bar{\alpha}(\mu)$. The screening mass and the free 
energy ${\cal F}^{(4)}$ for the $n=4$ truncation can be obtained simply by 
substituting $\bar{\alpha}\to 1/(1/\bar{\alpha}+L+\gamma)$ into 
(\ref{quadra-sol}) and (\ref{free:sqrt}). The resulting expressions are 
completely independent of $\mu$. 

The thermodynamic functions defined by truncating the $m/T$ expansion 
converge much more quickly than the weak-coupling expansion. This is 
illustrated in Fig.~3b, which shows the free energy divided by the ideal 
gas as a function of $\bar{g}(2\pi T)$. The heavy solid line is the exact 
2-loop $\Phi$-derivable result. The slighter solid lines are the results 
from truncating the thermodynamic potential after terms of order $g^n$, 
$n=2,3,4,6$, with $m$ treated as order $g$. For $n=2,3$, we set $\mu=2\pi T$, 
but beginning at order $g^4$, the result is completely independent of $\mu$. 
The results for $n=4,6$ are so close to the exact result that they cannot be 
distinguished from it in the Figure. Thus the convergence of the $m/T$ 
expansion is remarkably fast.

\subsection {Solution in terms of True Coupling Constant}

In the 3-loop $\Phi$-derivable approximation, we will not have the luxury 
of being able to calculate the thermodynamic potential as explicitly as 
in (\ref{thpot2:f}). The best we will be able to do is calculate the 
sum-integrals as truncated expansions in $m/T$. We will also be unable to 
absorb all the ultraviolet divergences in the free energy into a renormalized 
coupling constant $\bar{\alpha}(\mu)$ that runs differently from the true 
coupling constant $\alpha(\mu)$. We will therefore repeat the solution to 
the 2-loop $\Phi$-derivable approximation in a manner that is as parallel 
as possible to the method we will use for the 3-loop case. 

Our goal will be to construct a thermodynamic potential that depends on the 
true coupling constant $\alpha(\mu)$ defined by (\ref{bare-ren}) and is 
finite to as high an order in $g$ as is possible. We have the freedom to 
remove ultraviolet divergences from the thermodynamic potential 
(\ref{thpot2:1}) without changing the free energy by adding multiples of 
$(m^2-G)^2$, where $G$ is the right side of the gap equation (\ref{gap2:1}). 
We can cancel the divergences proportional to $g_0^2/\epsilon^2$ in the last 
term of (\ref{thpot2:1}) by subtracting $(m^2-G)^2/(2(g_0\mu^{-\epsilon})^2)$. 
The resulting thermodynamic potential is 
\begin{equation}
\mu^{2\epsilon}\Omega(m) = {1\over 2}\sumint_P\log(P^2+m^2)-{m^4\over 2
(g_0\mu^{-\epsilon})^2}\,.
\end{equation}
Expanding in powers of $m/T$, we get
\begin{equation}
{\mu^{2\epsilon}\Omega(m)\over {\cal F}_{\rm ideal}} = 1-15\left\{ \hat{m}^2
-4\hat{m}^3-3\left({1\over \alpha_0\mu^{-2\epsilon}}+{1\over 2\epsilon}
+L+\gamma\right)\hat{m}^4\right\} +{\cal O}(\hat{m}^6)\,.
\label{thpot:g}
\end{equation}
If we substitute the expression (\ref{bare-ren}) for the bare coupling 
constant in terms of the true renormalized coupling constant $\alpha(\mu)$, 
we find that there are still ultraviolet divergences proportional to 
$\hat{m}^4$:
\begin{equation}
{\mu^{2\epsilon}\Omega(m)\over {\cal F}_{\rm ideal}} = 1-15\left\{ \hat{m}^2
-4\hat{m}^3-3\left[{1\over \alpha}+\left(-{1\over \epsilon}+L+\gamma\right)
-{17\over 12\epsilon}\alpha\right]\hat{m}^4\right\} +{\cal O}(g^6)\,.
\end{equation}
The closest we can come to constructing a finite thermodynamic potential 
that is a function of the true coupling constant is to truncate after the 
terms of order $g^3$:   
\begin{equation}
{1\over 15}\left({\Omega(m)\over {\cal F}_{\rm ideal}}-1\right) = 
{3\over \alpha}\hat{m}^4 -\hat{m}^2+4\hat{m}^3.
\end{equation}
The corresponding gap equation is
\begin{equation}
\hat m^2 = {\alpha\over 6}\left[ 1-6\hat m\right]\,.
\label{simplegap}
\end{equation}
Note that these expressions differ from those obtained by truncating 
(\ref{free:n=6}) and (\ref{gap:N6}), because they involve the true coupling 
constant $\alpha$ instead $\bar \alpha$. The solution to the gap equation 
(\ref{simplegap}) is obtained by substituting $\alpha$ for $\bar \alpha$ in 
(\ref{quadra-sol}). The resulting expression for the free energy is obtained 
by substituting $\alpha$ for $\bar \alpha$ in (\ref{free:sqrt}).

\section{Three-Loop $\Phi$-derivable approximation}
\label{Three-loop}

In the three-loop $\Phi$-derivable approximation, the solution to the 
variational equation (\ref{vary:2}) for the self-energy $\Pi(P)$ is a 
nontrivial function of the momentum $P$. The thermodynamic potential in 
the 3-loop $\Phi$-derivable approximation is 
\begin{eqnarray}
\mu^{2 \epsilon} \Omega &=& 
{1 \over 2} \sumint_P \log (P^2 + \Pi(P))
- {1 \over 2} \sumint_P {\Pi(P) \over P^2 + \Pi(P)}
\nonumber
\\
&& 
+ {1 \over 8} (g_0 \mu^{-\epsilon})^2
        \left( \sumint_P {1 \over P^2 + \Pi(P)} \right)^2 
- {1 \over 48} (g_0 \mu^{-\epsilon})^4 \Im _{\rm ball} \,,
\label{thpot3:1}
\end{eqnarray}
where $\Im _{\rm ball}$ is the basketball sum-integral:
\begin{eqnarray}
\Im _{\rm ball} &=& 
\sumint_{PQR} {1 \over P^2 + \Pi(P)} {1 \over Q^2 + \Pi(Q)} 
	{1 \over R^2 + \Pi(R)} {1 \over S^2 + \Pi(S)} \,,
\label{IBall}
\end{eqnarray}
where $S = -(P+Q+R)$.  
The variational equation obtained by varying (\ref{thpot3:1})
with respect to $\Pi(P)$ is
\begin{eqnarray}
\Pi(P) &=& 
{1 \over 2} (g_0 \mu^{-\epsilon})^2 \sumint_Q {1 \over Q^2 + \Pi(Q)} 
- {1 \over 6} (g_0 \mu^{-\epsilon})^4 \Im _{\rm sun}(P) \,,
\label{vary3:1}
\end{eqnarray}
where $\Im _{\rm sun}(P)$ is the sunset sum-integral:
\begin{eqnarray}
\Im _{\rm sun}(P) &=& 
\sumint_{QR} {1 \over Q^2 + \Pi(Q)} {1 \over R^2 + \Pi(R)} 
	{1 \over S^2 + \Pi(S)} \,.
\label{ISun}
\end{eqnarray}
The three-loop $\Phi$-derivable approximation to the free
energy is obtained by solving (\ref{vary3:1}) for $\Pi(P)$ and inserting 
the solution into the thermodynamic potential (\ref{thpot3:1}).

\subsection {Mass Variable}

Our strategy for solving the 3-loop $\Phi$-derivable approximation is 
to introduce a mass variable $m$ which is of order 
$g T$ in the weak-coupling limit and then calculate the sum-integrals 
in (\ref{thpot3:1}) and (\ref{vary3:1}) as double expansions in 
$\alpha_0$ and $m/T$. We will find that the gap equation (\ref{vary3:1}) 
for $\Pi(P)$ has a recursive structure that allows us to solve for its 
dependence on $P$. Inserting the solution into (\ref{thpot3:1}), we can 
reduce the thermodynamic potential $\Omega(m)$ to a double expansion in 
$\alpha_0$ and $m/T$. 

The simplest choice for the variational parameter conceptually is the 
self-energy at zero external momentum $\Pi(0)$. This is simply the value 
of the variational function $\Pi(P)$ at a particular point in momentum space. 
One of the disadvantages of $\Pi(0)$ as a variational parameter is that it is 
ultraviolet divergent. Renormalization theory guarantees that in 
conventional perturbation theory the ultraviolet divergences in the exact 
propagator $1/(P^2+\Pi(P))$ are canceled order-by-order in the 
renormalized coupling constant $\alpha$ by multiplying 
by $Z^{-1}$, where $Z$ is the wavefunction renormalization constant. Thus 
$Z\Pi(0)$ should be finite order-by-order in $\alpha$. Since $Z$ is divergent 
at order $\alpha^2$ and $\Pi(0)$ begins at order $\alpha$, $\Pi(0)$  must 
have ultraviolet divergences at order $\alpha^3$. Now the $\Phi$-derivable 
approximation involves a truncation of perturbation theory, so multiplicative 
renormalizability is guaranteed only up to the truncation error. Since the 
three-loop $\Phi$-derivable approximation does not include all three-loop 
self-energy diagrams, the truncation error for $Z\Pi(0)$ is of order 
$\alpha^3$. If we use $\Pi(0)$ as a variational parameter, ultraviolet 
divergences in the solution to the gap equation will come partly from true 
divergences of $\Pi(0)$ and partly from the failure of multiplicative 
renormalizability in the 3-loop $\Phi$-derivable approximation. 

A choice for the variational parameter with more convenient ultraviolet 
properties is the screening mass defined by
\begin{equation}
\Pi (0, {\bf p}) \to m_s^2 \qquad {\rm as} \; p^2 \to - m_s^2.
\label{screen:1}
\end{equation}
Renormalization theory guarantees that in the conventional perturbation 
theory $m_s^2$ is finite order-by-order in the renormalized coupling 
constant $\alpha$. Since the three-loop $\Phi$-derivable approximation 
does not include all three-loop self-energy diagrams, $m_s^2$ is guaranteed 
to be finite only up to corrections of order $\alpha^3$. But any divergences 
at this order will come from the failure of multiplicative renormalizability 
in the 3-loop $\Phi$-derivable approximation. 

In the three-loop $\Phi$-derivable approximation, the screening mass $m_s$ 
defined by (\ref{screen:1}) is the solution to the gap equation
\begin{eqnarray}
m^2 &=& 
{1 \over 2} (g_0 \mu^{-\epsilon})^2 \sumint_Q {1 \over Q^2 + \Pi(Q)} 
- {1 \over 6} (g_0 \mu^{-\epsilon})^4 
	\Im _{\rm screen} \,,
\label{screen:2}
\end{eqnarray}
where we have introduced the short-hand 
\begin{equation}
\Im _{\rm screen}\equiv \Im _{\rm sun}(0,{\bf p})\bigg|_{p=im}\,.
\end{equation} 
Since the sum-integrals in (\ref{screen:2}) can be expressed as expansions 
in $\Pi(0)/T$, this equation defines $m^2$ as a function of $\Pi(0)$. We 
will find it more convenient to calculate the sum-integrals as expansions 
in $m/T$. Subtracting (\ref{screen:2}) from (\ref{vary3:1}), the variational 
equation reduces to 
\begin{eqnarray}
\Pi(P) &=& m^2
- {1 \over 6} (g_0 \mu^{-\epsilon})^4 
\left[ \Im _{\rm sun}(P) 
	- \Im _{\rm screen} \right] \,.
\label{vary3:2}
\end{eqnarray}
If we set $P=0$ and evaluate the sunset sum-integrals on the right-side as 
expansions in $m/T$, we get an equation for $\Pi(0)$ as a function of $m$. 
This guarantees that the mass variable $m$ defined by (\ref{screen:1}) can 
be used as a variational parameter in place of $\Pi(0)$. The solution to 
the gap equation (\ref{screen:2}) for $m$ will be the $\Phi$-derivable 
approximation to the screening mass $m_s$.

\subsection {Solution to the Variational Equation}

In the variational equation (\ref{vary3:2}) for the self-energy 
$\Pi(P)$, the sum-integrals extend over all momenta.  
There are two important momentum scales:  
the {\it hard} scale $2 \pi T$ and the {\it soft} scale $m$. 
The hard region for the momentum $P= (2 \pi nT, {\bf p})$ 
includes $n \neq 0$ for all ${\bf p}$ 
and also $n = 0$ with ${\bf p}$ of order $T$. 
The soft region is $n = 0$ and ${\bf p}$ of order $m$.  
We will solve the variational equation in the two momentum regions separately.

In order to solve the variational equation for $\Pi (P)$, we will assume
that the scales $m$ and $2 \pi T$ can be separated.  This will allow $\Pi
(P)$ to be expanded in powers of $g_0 \mu^{- \epsilon}$ and $m/T$.  
In the variational equation (\ref{vary3:2}), the bare coupling constant 
appears only in the combination $(g_0 \mu^{- \epsilon})^4$.
We therefore assume that the self-energy $\Pi(P)$ can be expanded 
in powers of $(g_0 \mu^{- \epsilon})^4$, 
with coefficients that are functions of $P$, $m$ and $T$.
These functions in turn can be expanded in powers of $m/T$.
For hard momentum $P$, we write the expansion in the form
\begin{eqnarray}
\Pi (P) &=& m^2 
\;+\; (g_0 \mu^{- \epsilon})^4 \left[ \Pi_{4,0}(P) +  \Pi_{4,1}(P) 
					+ \Pi_{4,2}(P) + \Pi_{4,3}(P)
+ \ldots \right]
\nonumber
\\
&& \hspace{1cm}
\;+\; (g_0 \mu^{- \epsilon})^8 \left[ \Pi_{8,-2}(P) + \Pi_{8,-1}(P) + 
\ldots \right]
\;+\; \ldots \qquad {\rm for \; hard} \; P \,,
\label{Pi-hard}
\end{eqnarray}
where $\Pi_{n,k}(P)$ is of order $T^2(m/T)^k$ when $P$ is of order $T$.
We have anticipated the solution to the variational equation 
by writing explicitly all terms that contribute through order $g^7 T^2$. 
For soft momentum $P = (0, {\bf p})$, we write 
the self-energy as $\Pi (0, {\bf p}) = m^2 + \sigma(p)$. 
The $m/T$ expansion of the function $\sigma(p)$ has the form
\begin{eqnarray}
\sigma(p) &=& 
(g_0 \mu^{- \epsilon})^4 \left[ \sigma_{4,-2}(p) + \sigma_{4,0}(p) +
\sigma_{4,1}(p) + \ldots \right]
\nonumber
\\
&& \hspace{1cm}
\;+\; (g_0 \mu^{- \epsilon})^8 \left[ \sigma_{8,-4}(p) + \ldots \right]
\;+\; \dots \; \hspace{1cm} {\rm for \; soft} \; p \,,
\label{Pi-soft}
\end{eqnarray}
where $\sigma_{n,k}(p)$ is of order $m^2(m/T)^k$ when $p$ is of order $m$.
We have anticipated the solution to the variational equation 
by writing explicitly all terms that contribute through order $g^5 m^2$. 
Upon inserting the expansions (\ref{Pi-hard}) and (\ref{Pi-soft}) into the
variational equation (\ref{vary3:2}) and expanding systematically in
powers of $g_0 \mu^{-\epsilon}$ and $m /T$, we will find that it has a
recursive structure that either determines $\Pi_{n,k}(P)$ and 
$\sigma_{n,k}(p)$ or allows them to be expressed in terms of lower order 
functions.

The $m/T$ expansions for the sunset sum-integral $\Im _{\rm sun}(P)$ are 
derived in Appendix \ref{app:movert}. For hard momentum $P$, the $m/T$ 
expansion including all terms up to errors of order $g^4T^2$ is the sum of 
(\ref{ISunh:hh}), (\ref{ISunh:hs}), and (\ref{ISunh:ss}):
\begin{eqnarray}
\Im _{\rm sun} (P) &=& \sumint_{QR} {1 \over Q^2 R^2 
(P+Q+R)^2} + 3 T I_1 \sumint_Q {1 \over Q^2 (P + Q)^2} \nonumber
\\
&& + 3 T^2 I_1^2 {1 \over P^2} -3m^2\sumint_{QR}{1\over (Q^2)^2 R^2 
(P+Q+R)^2}   \nonumber
\\
&&- 3 T m^2 I_1 \sumint_Q \left( {1 \over (Q^2)^2 (P + Q)^2} +
{(4/d){\bf q}^2 \over (Q^2)^3 (P + Q)^2}\right)
\nonumber
\\
&& - 3 (g_0 \mu^{-\epsilon})^4 T 
	\int_{\bf r} {\sigma_{4,-2}(r) \over (r^2 + m^2)^2} 
	\sumint_Q {1 \over Q^2 (P+Q)^2} + {\cal O} (g^4 T^2) \,.
\label{sunexp:h}
\end{eqnarray}
The momentum integral $I_1$ is proportional to $m^{1-2\epsilon}$ and is 
given in (\ref{bi1}). We will find that the momentum integral involving 
$\sigma_{4,-2}$ is proportional to $m^{-3-4\epsilon}$. For soft momentum 
$P=(0,{\bf p})$, the sunset sum-integral, including all terms up to errors 
of order $g^4T^2$, is the sum of (\ref{ISuns:hh}), (\ref{ISuns:hs}), and 
(\ref{ISuns:ss}):
\begin{eqnarray}
\Im _{\rm sun}(0, {\bf p}) &=& T^2I_{\rm sun}(p) + 3T I_1 \sumint_Q 
{1 \over (Q^2)^2} \nonumber
\\
&&- (p^2 + 3m^2) \sumint_{QR} 
{1 \over (Q^2)^2 R^2 (Q+R)^2} + {4 \over d} p^2 \sumint_{QR} 
{ {\bf q}^2 \over (Q^2)^3 R^2 (Q+R)^2} \nonumber
\\
&& - 3T (p^2 + m^2) I_1 \sumint_Q {1 \over (Q^2)^3} 
+ {12 \over d} T (p^2 - m^2 ) I_1 \sumint_Q{{\bf q}^2 \over (Q^2)^4} \nonumber
\\
&& - 3 (g_0 \mu^{- \epsilon})^4 \left[ 
T \int_{\bf r} {\sigma_{4,-2} (r) \over (r^2 + m^2)^2} \sumint_Q 
{1 \over (Q^2)^2} +  T^2 \int_{\bf q} {\sigma_{4,-2} (q) \over (q^2 + m^2)^2}
I_{\rm bub}(|{\bf p}+{\bf q}|) \right] \nonumber
\\
&& + {\cal O} (g^4 T^2) \,.
\label{sunexp:s}
\end{eqnarray}
The sunset integral $I_{\rm sun}(p)$ is defined in (\ref{Isun}) and scales 
like $m^{-4-2\epsilon}$ when $p$ is of order $m$. The bubble integral 
$I_{\rm bub}(p)$ is defined in (\ref{Ibub}) and scales like 
$m^{-1-2\epsilon}$ when $p$ is of order $m$.

We proceed to solve the variational equation (\ref{vary3:2}) for $\Pi(P)$. 
We first consider the equation for hard momentum $P$.  On the left side 
of (\ref{vary3:2}), we insert the expansion (\ref{Pi-hard}) for $\Pi (P)$. 
On the right side of (\ref{vary3:2}), we insert the $m/T$ expansions 
(\ref{sunexp:h}) and (\ref{sunexp:s}) of the sunset sum-integrals. 
Matching coefficients of $(g_0 \mu^{ - \epsilon})^{4n}$ on both sides of
the equation and then identifying the terms in the $m/T$ expansions, we find
\begin{eqnarray}
\Pi_{4,0} (P)  &=&  
- {1 \over 6} \sumint_{QR} {1 \over Q^2 R^2 (P+Q+R)^2} 
	+{1\over 6} T^2 I_{\rm sun} (im)\,,
\label{Pi40}
\\
\Pi_{4,1} (P)  &=& 
- {1 \over 2} T I_1 \sumint_Q \left ( {1 \over Q^2 (P+Q)^2} 
					- {1 \over (Q^2)^2} \right)\,,
\label{Pi41}
\\
\Pi_{4,2} (P)  &=&  - {1 \over 2} T^2 I_1^2 {1 \over P^2} +
{1 \over 2} m^2 \sumint_{QR} {1 \over (Q^2)^2 R^2 (P+Q+R)^2}
\nonumber
\\ 
&& - {1\over3} m^2 \sumint_{QR}\left( {1 \over (Q^2)^2 R^2 (Q+R)^2} 
+ {(2/d){\bf q}^2 \over (Q^2)^3 R^2 (Q + R)^2} \right)\,,
\\
\Pi_{8,-2}  &=&  
-{1 \over 2} T^2 \int_{\bf q} 
	{\sigma_{4,-2} (q) \over (q^2 + m)^2} I_{\rm bub}(|{\bf k}+{\bf q}|)
\bigg|_{k=im}\,.
\label{Pi8-2}
\end{eqnarray}
We next consider the variational equation for soft momentum $P=(0,{\bf p})$.  
On the left side of (\ref {vary3:2}), we insert the expansion
(\ref{Pi-soft}) for $\Pi (0, {\bf p})$. On the right side of (\ref{vary3:2}), 
we insert the $m/T$ expansions (\ref{sunexp:s}) of the sunset sum-integral 
at soft momentum. Matching coefficients of $(g_0 \mu^{-\epsilon})^{4n}$ on 
both sides of the equation and then identifying the terms in the $m/T$ 
expansions, we find
\begin{eqnarray}
\sigma_{4,-2} (p)  &=&  
- {1\over 6}T^2 \left[ I_{\rm sun}(p) - I_{\rm sun}(im) \right] \,,
\label{sigma4-2}
\\
\sigma_{4,0} (p)  &=& 
{1 \over 6}(p^2+m^2) \sumint_{QR} \left( {1 \over (Q^2)^2 R^2 (Q+R)^2} 
		- {(4/d) {\bf q}^2 \over (Q^2)^3 R^2 (Q+R)^2} \right) \,,
\\
\sigma_{8,-4} (p)  &=&  
{1\over 2}T^2 \int_{\bf q} {\sigma_{4,-2} (q) \over (q^2 + m^2)^2} 
			I_{\rm bub}(|{\bf p}+{\bf q}|) +\Pi_{8,-2}\,.
\label{sigma8-4}
\end{eqnarray}
Notice that the matching equations in (\ref{Pi40})--(\ref{sigma8-4}) 
have a recursive structure.  The functions $\Pi_{4, 0}$,
$\Pi_{4, 1}$, $\Pi_{4, 2}$, $\sigma_{4, -2}$, and $\sigma_{4, 0}$
are given explicitly in terms of integrals and sum-integrals.  The
functions $\Pi_{8, -2}$ and $\sigma_{8, -4}$ are given in terms of
integrals of the function $\sigma_{4, -2} (q)$ which has already been
determined. The recursive structure of these equations continues at
higher order in $g$. Thus, by using this method, we can solve the variational 
equation for $\Pi(P)$ order-by-order in $g_0^4$ and $m/T$ for both hard $P$ 
and soft $P$. 

In order to calculate the thermodynamic potential, we will need to calculate 
the sum-integrals and integrals involving some of the functions 
$\Pi_{m,k}(P)$ and $\sigma_{m,k}(p)$. Using the solutions (\ref{Pi40}) and 
(\ref{Pi41}), we obtain the following sum-integrals:
\begin{eqnarray}
\sumint_P {\Pi_{4, 0}(P) \over (P^2)^2} & = & 
- {1 \over 6} \sumint_{PQR} {1 \over (P^2)^2 Q^2 R^2 (P+Q+R)^2} 
+ {1 \over 6} T^2 I_{\rm sun} (im) \sumint_P {1 \over (P^2)^2} \,,
\label{Pi40:P4}
\\
\sumint_P {\Pi_{4, 1}(P) \over (P^2)^2} & = & 
- {1 \over 2} T I_1 
\left[ \sumint_{PQ} {1 \over (P^2)^2 Q^2 (P+Q)^2} 
		- \left( \sumint_P {1 \over (P^2)^2} \right)^2 \right] \,.
\label{Pi41:P4}
\end{eqnarray}
Using the solutions (\ref{sigma4-2})-(\ref{sigma8-4}), we obtain the 
following integrals:
\begin{eqnarray}
T\int_{\bf p} {\sigma_{4, -2} (p) \over (p^2 + m^2)^2} & = & 
- {1 \over 6} T^3 \int_{\bf p} 
	{I_{\rm sun} (p) - I_{\rm sun} (im) \over (p^2 + m^2)^2} \,,
\label{sigma4-2:p2m2}
\\
T\int_{\bf p} {\sigma_{4, 0} (p) \over (p^2 + m^2)^2} & = &  
{1 \over 6} T I_1 \sumint_{PQ} 
	\left( {1 \over (P^2)^2 Q^2 (P+Q)^2} 
		- {(4/d) {\bf p}^2 \over (P^2)^3 Q^2 (P+Q)^2} \right) \,,
\\
T\int_{\bf p} {[\sigma_{4, -2} (p)]^2 \over (p^2 + m^2)^3} & = & 
{1 \over 36} T^5 \int_{\bf p} 
	{[I_{\rm sun} (p) - I_{\rm sun} (im) ]^2 \over (p^2 + m^2)^3} \,,
\\
T\int_{\bf p} {\sigma_{8, -4} (p) \over (p^2 + m^2)^2} & = & {1\over 36}T^5
\int_{\bf p} {I_{\rm sun} (p) - I_{\rm sun} (im) \over (p^2 + m^2)^2}
{d\over dm^2}I_{\rm sun}(p) + T I_2\Pi_{8, -2}\, .
\label{sigma8-4:p2m2}
\end{eqnarray}

\subsection {Gap Equation}

Upon inserting our solutions to the variational equation into 
(\ref{thpot3:1}) and expanding it in powers of $m/T$, we will obtain a 
thermodynamic potential $\Omega_0(m)$ that depends on a single variational 
parameter $m$. This function will contain severe ultraviolet divergences. 
Our goal will be to construct a thermodynamic potential that is a finite 
function of $m$ but has the same variational minimum. To accomplish this, we 
will add to $\Omega_0(m)$ a quantity that includes a factor of $(m^2-G)^2$, 
where $G$ is the right side of the gap equation (\ref{screen:2}). We will 
therefore need this gap equation in the form of an explicit equation for $m$. 

To make the gap equation (\ref{screen:2}) explicit, we must expand the 
sum-integrals on the right side in powers of $m/T$. We will keep all finite 
terms through order $g^5$ and all divergent terms through order $g^7$. Our 
first step will be to expand them in powers of the self-energy functions 
$\Pi_{n,k}(P)$ and $\sigma_{n,k}(p)$. Each of the sum-integrals in 
(\ref{thpot:e1}) can be expressed as a sum of the corresponding massive 
sum-integral, sum-integrals over $P$ involving the functions $\Pi_{n,k}(P)$, 
and integrals over ${\bf p}$ involving the functions $\sigma_{n,k}(p)$. 
The terms that contribute to the gap equation through order $g^7$ are
\begin{eqnarray}
&&\sumint_P{1\over P^2+\Pi(P)}={\cal I}_1-(g_0\mu^{-\epsilon})^4 \left[
T\int_{\bf q}{\sigma_{4,-2}(q)+\sigma_{4,0}(q)\over (q^2+m^2)^2}
+\sumint_Q{\Pi_{4,0}(Q)+\Pi_{4,1}(Q)\over (Q^2)^2} \right] \nonumber
\\
&&\hspace{3.5cm}
-(g_0\mu^{-\epsilon})^8\int_{\bf q}\left( {\sigma_{8,-4}(q)\over 
(q^2+m^2)^2}- {\sigma_{4,-2}^2(q)\over (q^2+m^2)^3} \right)
+{\cal O}(g^6)\,,
\label{tadpol}
\\
&&\hspace{1.5cm}
\Im_{\rm screen}={\cal I}_{\rm screen}-3(g_0\mu^{-\epsilon})^4\left[
T^2\int_{\bf q}{\sigma_{4,-2}(q)\over 
(q^2+m^2)^2}I_{\rm bub}(|{\bf k}+{\bf q}|)\bigg|_{k=im}\right. \nonumber
\\
&&\hspace{6.5cm}\left.
+T\int_{\bf q}{\sigma_{4,-2}(q)\over (q^2+m^2)^2}\sumint_Q{1\over (Q^2)^2}
\right] +{\cal O}(g^4)\,, 
\label{IIscreen}
\end{eqnarray}
The resulting form of the gap equation, including all term through order 
$g^7T^2$, is
\begin{eqnarray}
m^2 &=& (g_0\mu^{-\epsilon})^2\left[ {1\over 2}{\cal I}_1 \right]
+ (g_0\mu^{-\epsilon})^4\left[ -{1\over 6}{\cal I}_{\rm screen} \right]
\nonumber
\\
&& + (g_0 \mu^{- \epsilon})^6 
\left\{ - {1\over 2}T \int_{\bf p} {\sigma_{4,-2}(p) + \sigma_{4,0}(p) 
				\over (p^2 + m^2)^2} 
	- {1\over 2}\sumint_P {\Pi_{4,0}(P) + \Pi_{4,1}(P) \over (P^2)^2} 
\right\} \nonumber
\\
&&+ (g_0 \mu^{- \epsilon} )^8 
\left\{ {1\over 2} T^2 \int_{\bf q}{\sigma_{4, -2} (q) \over (q^2 + m^2)^2} 
I_{\rm bub}(|{\bf p}+{\bf q}|)\bigg|_{p=im}
+{1\over 2}T \int_{\bf q} {\sigma_{4,-2}(q) \over (q^2 + m^2)^2}
	\sumint_P {1 \over (P^2)^2} \right \} \nonumber
\\
&& + (g_0 \mu^{- \epsilon})^{10} 
\left\{ -{1\over 2}T \int_{\bf q} \left(  {\sigma_{8,-4} (q) 
\over (q^2 + m^2)^2} - {\sigma_{4, -2}^2 (q) \over (q^2 + m^2)^3} \right) 
\right\} + {\cal O} (g^8 T^2) \,.
\label{gap3:2}
\end{eqnarray}

The $m/T$ expansions of the massive sum-integrals ${\cal I}_1$ and 
${\cal I}_{\rm screen}$ are given in  (\ref{Imass:1}) and (\ref{Iscreen:m2}). 
The only momentum integrals involving self-energy functions that contribute 
to the finite terms through order $g^5T^2$ or to divergent terms through 
order $g^7T^2$ are
\begin{eqnarray}
T\int_{\bf q}{\sigma_{4,-2}(q)\over (q^2+m^2)^2} &=& {T^2\over (4\pi)^4}
{1-\log 2\over 6}{1\over \hat{m}}\,, 
\label{sigma4-2:d2}
\\
T\int_{\bf q}{\sigma_{4,0}(q)\over (q^2+m^2)^2} &=& {T^2\over (4\pi)^4}
\left[ -{1\over 48\epsilon}\right]\hat{m}\,.
\end{eqnarray} 
The sum-integrals involving self-energy functions are
\begin{eqnarray}
\sumint_P {\Pi_{4, 0}(P) \over (P^2)^2} & = & {T^2\over 48(4\pi)^4}
\left({\hat{\mu}\over 2}\right)^{6\epsilon}\left[{1\over \epsilon^2}
+\left(-8\log \hat{m} -{17\over 6}-16\log 2-2{\zeta'(-1)\over \zeta(-1)}
\right){1\over \epsilon}\right]\,,
\\
\sumint_P {\Pi_{4, 1}(P) \over (P^2)^2} & = & -{T^2\over 8(4\pi)^4}
\left({\hat{\mu}\over 2}\right)^{6\epsilon}\left[{1\over \epsilon^2}
+\left(-2\log \hat{m} +1+4\gamma\right){1\over \epsilon}\right]\hat{m} \,. 
\end{eqnarray}
Keeping terms of order $\epsilon$ through order $g^3$, finite terms through 
order $g^5$, and divergent terms through order $g^7$, the gap equation 
(\ref{gap3:2}) becomes
\begin{eqnarray}
\hat{m}^2 &=& 
(\alpha_0 \mu^{- 2\epsilon})\left({ \hat{\mu}\over 2}\right)^{2\epsilon} 
\left\{ {1\over 6}\left[ 1+\left(2+2{\zeta'(-1)\over \zeta(-1)}\right)
\epsilon  \right]-\left[ 1+\left( -2\log \hat{m} +2 \right)\epsilon
\right]\hat{m} \right.\nonumber
\\
&&\hspace{3cm} \left.
-{1\over 2}\left[{1\over \epsilon}+2\gamma\right]\hat{m}^2\right\} \nonumber
\\
&&+(\alpha_0 \mu^{- 2\epsilon})^2\left({ \hat{\mu}\over 2}\right)^{4\epsilon} 
\left\{ -{1\over 6}\left[{1\over \epsilon}+\left(-4\log \hat{m}
+6-8\log 2 \right)\right]\right. \nonumber
\\
&&\hspace{3cm} 
+\left[ {1\over \epsilon}+\left(-2\log \hat{m}
+2+2\gamma\right)\right]\hat{m} \nonumber
\\
&&\hspace{3cm} \left.
+{1\over 4}\left[ {1\over \epsilon^2}+\left( {5\over 6}+4\gamma\right)
{1\over \epsilon}\right]\hat{m}^2\right\} \nonumber
\\
&&+(\alpha_0 \mu^{- 2\epsilon})^3\left({ \hat{\mu}\over 2}\right)^{6\epsilon}
\left\{ -{1-\log 2\over 3}{1\over \hat{m}} \right. \nonumber
\\
&&\hspace{3cm} 
-{1\over 24}\left[ {1\over \epsilon^2}+\left( -8\log \hat{m}-
{17\over 6}-16\log 2-2{\zeta'(-1)\over \zeta(-1)}
\right){1\over \epsilon}\right] \nonumber
\\
&&\hspace{3cm} \left.
+{1\over 4}\left[ {1\over \epsilon^2}+\left( -2\log \hat{m}+{7\over 6}
+4\gamma\right){1\over \epsilon}\right]\hat{m} \right\} \nonumber
\\
&&+(\alpha_0 \mu^{- 2\epsilon})^4\left({ \hat{\mu}\over 2}\right)^{8\epsilon}
\left\{ {1-\log 2\over 3\epsilon}{1\over \hat{m}} \right\}  \,,
\label{gap:g5}
\end{eqnarray}

\subsection {Thermodynamic Potential}

The 3-loop $\Phi$-derivable approximation to the thermodynamic potential 
$\Omega_0[D]$ is given in (\ref{thpot3:1}) as a functional of the self-energy 
$\Pi(P)$. The sum-integrals can be expanded in powers of $m/T$ using the 
methods of Appendix~\ref{app:movert}. The function $\Pi(P)$ can then be 
eliminated in favor of the mass variable $m$ by using the solution to the 
variational equation in subsection \ref{Three-loop}.B. This reduces the 
thermodynamic potential to a function $\Omega_0(m)$ of a single variational 
parameter $m$. This function contains many ultraviolet divergences. Some of 
them are eliminated when $\Omega_0(m)$ is evaluated at the solution of the 
gap equation. Some can be eliminated by renormalization of the coupling 
constant. There are other ultraviolet divergences that cannot be 
canceled and represent unavoidable ambiguities in the $\Phi$-derivable 
approximation. We wish to construct a thermodynamic potential that is finite 
to as high an order in $g$ as possible. 

Our first step in constructing a finite thermodynamic potential $\Omega(m)$ 
will be to cancel the most severe divergences that are eliminated 
by evaluating $\Omega_0(m)$ at the solution to the gap equation. These are the 
divergences proportional to $m^4/\epsilon$ and $g_0^2m^4/\epsilon^2$ that 
were already encountered in the 2-loop $\Phi$-derivable approximation. They 
can be canceled without changing the free energy by adding the term
\begin{equation}
-{1\over 2(g_0\mu^{-\epsilon})^2}\left[\Pi(0)-{1\over 2}(g_0\mu^{-\epsilon})
^2\sumint_Q{1\over Q^2+\Pi(Q)}+{1\over 6}(g_0\mu^{-\epsilon})^4
\Im _{\rm sun}(0)\right]^2 \,.
\end{equation} 
Since this is proportional to the square of the gap equation evaluated at 
$P=0$, it vanishes at the solution to the gap equation. Using (\ref{vary3:2}) 
to eliminate $\Pi(0)$ in favor of $m$, this term can be written 
\begin{equation}
-{1\over 2(g_0\mu^{-\epsilon})^2}\left[m^2-{1\over 2}(g_0\mu^{-\epsilon})
^2\sumint_Q{1\over Q^2+\Pi(Q)}+{1\over 6}(g_0\mu^{-\epsilon})^4
\Im _{\rm screen}\right]^2 \,.
\label{addterm}
\end{equation}
The thermodynamic potential $\Omega_1(m)$ defined by adding (\ref{addterm}) to 
(\ref{thpot3:1}) is
\begin{eqnarray}
\mu^{2 \epsilon} \Omega_1(m) &=& -{m^4\over 2(g_0\mu^{-\epsilon})^2} +
{1 \over 2} \sumint_P \left( \log (P^2 + \Pi(P))
- {\Pi(P)-m^2 \over P^2 + \Pi(P)} \right) \nonumber
\\
&&+(g_0\mu^{-\epsilon})^2\left[-{1\over 6}m^2\Im _{\rm screen}\right] \nonumber
\\
&&+(g_0\mu^{-\epsilon})^4\left[-{1\over 48}\Im _{\rm ball}+{1\over 12}
\Im _{\rm screen}\sumint_P {1 \over P^2 + \Pi(P)}\right] \nonumber
\\
&&+(g_0\mu^{-\epsilon})^6\left[-{1\over 72}\Im _{\rm screen}^2\right] \,.
\label{thpot:e1}
\end{eqnarray}

We want to expand the thermodynamic potential (\ref{thpot:e1}) in powers of 
$m/T$, keeping all finite terms through order $g^5$ and all divergent terms 
through order $g^7$. Our first step will be to expand in powers of the 
self-energy functions $\Pi_{n,k}(P)$ and $\sigma_{n,k}(p)$. Each of the 
sum-integrals in (\ref{thpot:e1}) can be expressed as a sum of the 
corresponding massive sum-integral, sum-integrals over $P$ involving the 
functions $\Pi_{n,k}(P)$, and integrals over ${\bf p}$ involving the 
functions $\sigma_{n,k}(p)$. For the tadpole sum-integral and for 
$\Im _{\rm screen}$, these expressions are given in (\ref{tadpol}) and 
(\ref{IIscreen}). For the other two sum-integrals in (\ref{thpot:e1}), the 
terms that contribute to the thermodynamic potential through order $g^7$ are
\begin{eqnarray}
&&\hspace{2cm}
\sumint_P\left(\log(P^2+\Pi(P))-{\Pi(P)-m^2\over P^2+\Pi(P)}\right) 
\nonumber
\\
&&\hspace {2.7cm}
=-{\cal I}_0'+{1\over 2}(g_0\mu^{-\epsilon})^8T\int_{\bf q}
{\sigma_{4,-2}^2(q)\over (q^2+m^2)^2}+{\cal O}(g^8T^4)\,, 
\\
&&\hspace{1.8cm}
\Im _{\rm ball}={\cal I}_{\rm ball}-4(g_0\mu^{-\epsilon})^4T^3
\int_{\bf q}{\sigma_{4,-2}(q)I_{\rm sun}(q)\over (q^2+m^2)^2}
+{\cal O}(g^4T^4)\,.
\end{eqnarray}
The resulting expression for the thermodynamic potential, including all 
terms through order $g^7$, is 
\begin{eqnarray}
\mu^{2 \epsilon} \Omega_1(m) &=& -{m^4\over 2(g_0\mu^{-\epsilon})^2} +
{1\over 2}(-{\cal I}_0')
+(g_0\mu^{-\epsilon})^2\left[-{1\over 6}m^2{\cal I}_{\rm screen}\right] 
\nonumber
\\
&&+(g_0\mu^{-\epsilon})^4\left[-{1\over 48}{\cal I}_{\rm ball}+{1\over 12}
{\cal I}_{\rm screen}{\cal I}_1\right] \nonumber
\\
&&+(g_0\mu^{-\epsilon})^6\left[-{1\over 72}{\cal I}_{\rm screen}^2
+{1\over 2}m^2T^2\int_{\bf q}{\sigma_{4,-2}(q)\over 
(q^2+m^2)^2}I_{\rm bub}(|{\bf k}+{\bf q}|)\bigg|_{k=im} \right. \nonumber
\\
&&\hspace{2cm} \left.
+{1\over 2}m^2T \sumint_P {1\over (P^2)^2}
\int_{\bf q}{\sigma_{4,-2}(q)\over (q^2+m^2)^2} \right] 
\nonumber
\\
&&+(g_0\mu^{-\epsilon})^8\left[ -{1\over 4}{\cal I}_1T^2\int_{\bf q}
{\sigma_{4,-2}(q)\over (q^2+m^2)^2}I_{\rm bub}
(|{\bf k}+{\bf q}|)\bigg|_{k=im} \right. \nonumber
\\
&&\hspace{2cm} 
-{1\over 12}\left( {\cal I}_{\rm screen}+3{\cal I}_1\sumint_Q{1\over 
(Q^2)^2}\right) T\int_{\bf q}{\sigma_{4,-2}(q)\over 
(q^2+m^2)^2} \nonumber
\\
&&\hspace{2cm} \left.
+{1\over 4}T\int_{\bf q}{\sigma_{4,-2}^2(q)\over (q^2+m^2)^2}
+{1\over 12}T^3\int_{\bf q}{\sigma_{4,-2}(q)I_{\rm sun}(q)\over (q^2+m^2)^2}
\right] +{\cal O}(g^8T^4) \,.
\label{Omega1}
\end{eqnarray} 
The next step is to express the coefficient of each power of $g_0$ as an 
expansion in powers of $m/T$. The expansions of the sum-integrals 
${\cal I}_0'$, ${\cal I}_1$, ${\cal I}_{\rm screen}$, and 
${\cal I}_{\rm ball}$ are given in Appendix~\ref{app:mass-sumint}. The 
integrals involving the self-energy function $\sigma_{4,-2}(q)$ are given 
explicitly in (\ref{sigma4-2:p2m2})-(\ref{sigma8-4:p2m2}). The only integrals 
that contribute to the finite terms in (\ref{Omega1}) through order $g^5$ 
or to divergent terms through order $g^7$ are (\ref{sigma4-2:d2}) and 
\begin{equation}
T\int_{\bf q}{\sigma_{4,-2}(q)I_{\rm sun}(q)\over (q^2+m^2)^2} =
{T^2\over (4\pi)^6}{1-\log 2\over 24\epsilon}{1\over \hat{m}}\,.
\end{equation} 
The complete $m/T$ expansion of the thermodynamic potential, including all 
finite terms through order $g^5$ and all divergent terms through order 
$g^7$, is
\begin{eqnarray}
&& {1\over 15}\left({\mu^{2\epsilon}\Omega_1(m)\over {\cal F}_{\rm ideal}}
-1\right) = {3\over \alpha_0\mu^{-2\epsilon}}\hat{m}^4 
\nonumber
\\
&&\hspace{1cm} 
+\left({\hat{\mu}\over 2}\right)^{2\epsilon}\left[-\hat{m}^2+4\hat{m}^3
+{3\over 2}\left({1\over \epsilon}+2\gamma\right)
\hat{m}^4 \right] \nonumber 
\\
&&\hspace{1cm} 
+(\alpha_0\mu^{-2\epsilon})\left({\hat{\mu}\over 2}\right)^{4\epsilon}
\left\{ \left({1\over \epsilon}-4\log \hat{m}+6-8\log 2\right)
\hat{m}^2 \right. \nonumber
\\
&&\hspace{4cm}
-6\left({1\over \epsilon}-2\log \hat{m}+2+2\gamma\right)\hat{m}^3 \nonumber
\\
&&\hspace{4cm} \left.
-{3\over 2}\left[{1\over \epsilon^2}+\left({5\over 6}+4\gamma \right)
{1\over \epsilon}\right]\hat{m}^4 \right\} \nonumber
\\
&&\hspace{1cm} 
+(\alpha_0\mu^{-2\epsilon})^2\left({\hat{\mu}\over 2}\right)^{6\epsilon}
\left\{ -{1\over 12}\left[{1\over \epsilon}
-8\log \hat{m}+{149\over 15}-16\log 2 
-4{\zeta'(-1)\over \zeta(-1)}+2
{\zeta'(-3)\over \zeta(-3)}\right] \right. \nonumber
\\
&&\hspace{4cm}
+\left[{1\over \epsilon}-2\log \hat{m}+4-4\log 2+2\gamma+2
{\zeta'(-1)\over \zeta(-1)}\right]\hat{m} \nonumber
\\
&&\hspace{4cm}
+{1\over 2}\left[{1\over \epsilon^2}+\left(-4\log \hat{m} -6-8\log 2
+2\gamma\right) {1\over \epsilon}\right]\hat{m}^2 \nonumber
\\
&&\hspace{4cm}\left.
-3\left[{1\over \epsilon^2}+\left(-2\log \hat{m}+2+4\gamma\right)
{1\over \epsilon}\right]\hat{m}^3\right\} \nonumber
\\
&&\hspace{1cm}
+(\alpha_0\mu^{-2\epsilon})^3\left({\hat{\mu}\over 2}\right)^{8\epsilon}
\left\{ {1\over 12}\left[{1\over \epsilon^2}
+\left(-8\log \hat{m} +12-16\log 2\right){1\over \epsilon}\right] 
\right. \nonumber
\\
&&\hspace{4cm}\left.
-\left[{1\over \epsilon^2}+\left(-6\log \hat{m} +10-10\log 2+2\gamma\right)
{1\over \epsilon}\right]\hat{m}\right\} \nonumber
\\
&&\hspace{1cm} 
+(\alpha_0\mu^{-2\epsilon})^4\left\{ {1-\log 2\over 3\epsilon}
{1\over \hat{m}}\right\} \;. 
\end{eqnarray}  

This expression has ultraviolet divergences of order $g^4$ and higher. We 
have two ways to eliminate these divergences. One is to use (\ref{bare-ren}) 
to eliminate the bare coupling constant $\alpha_0$ in favor of the 
renormalized coupling constant $\alpha(\mu)$. The other is to add to the 
thermodynamic potential $\Omega(m)$ something proportional to the square of 
the gap equation. In particular, we can add a term such as (\ref{addterm}), 
with a coefficient proportional to any power of $(g_0\mu^{-\epsilon})^2$. We 
can define a thermodynamic potential $\Omega_2(m)$ in which the divergences 
are postponed to order $g^6$ as follows:
\begin{equation}
{\mu^{2\epsilon}\Omega_2(m)\over 15 {\cal F}_{\rm ideal}} = 
{\mu^{2\epsilon}\Omega_1(m)\over 15 {\cal F}_{\rm ideal}} + 
{3\over \epsilon}(\hat{m}^2-\hat{G})^2\left( {\mu\over \Lambda}\right)
^{2\epsilon}\,,
\label{addterm2}
\end{equation}
where $\hat{G}$ is the right side of the gap equation (\ref{gap:g5}). The 
factor of $\mu^{2\epsilon}$ in the term that is added guarantees that 
$\Omega_2(m)$ will be independent of $\mu$. The price that must be paid for 
the independence of $\mu$ is the introduction of another arbitrary scale 
$\Lambda$. After making the substitution (\ref{bare-ren}) and truncating 
after terms of $5^{\rm th}$ order in $\hat{m}$ or $g$, we obtain a finite 
thermodynamic potential
\begin{eqnarray}
{1\over 15}\left({\Omega(m)\over {\cal F}_{\rm ideal}} -1\right) &=& 
{3\hat{m}^4\over \alpha} +\left[ -\hat{m}^2+4\hat{m}^3+3\left(3L-2\ell+\gamma
\right)\hat{m}^4\right] \nonumber
\\
&&+\alpha\left[ 2\left( \ell-2\log \hat{m}+2-4\log 2 -{\zeta'(-1)\over 
\zeta(-1)}\right)\hat{m}^2-12\left(\ell+\gamma\right)\hat{m}^3\right] \nonumber
\\
&&+\alpha^2\left[ -{1\over 6}\left( \ell-4\log \hat{m}+{89\over 30}-
8\log 2-4{\zeta'(-1)\over \zeta(-1)}+
{\zeta'(-3)\over \zeta(-3)}\right) \right. \nonumber
\\
&&\hspace{1cm}  
+2\left( \ell-2\log 2+\gamma\right)\hat{m}\bigg] \,,
\label{thpot:fintrunc}
\end{eqnarray}
where $\ell=\log(\Lambda/4\pi T)$. This is our final result for the 3-loop 
$\Phi$-derivable thermodynamic potential.  
\begin{figure}[t]
\vspace*{-1cm}
\begin{tabular}{cc}
\epsfig{file=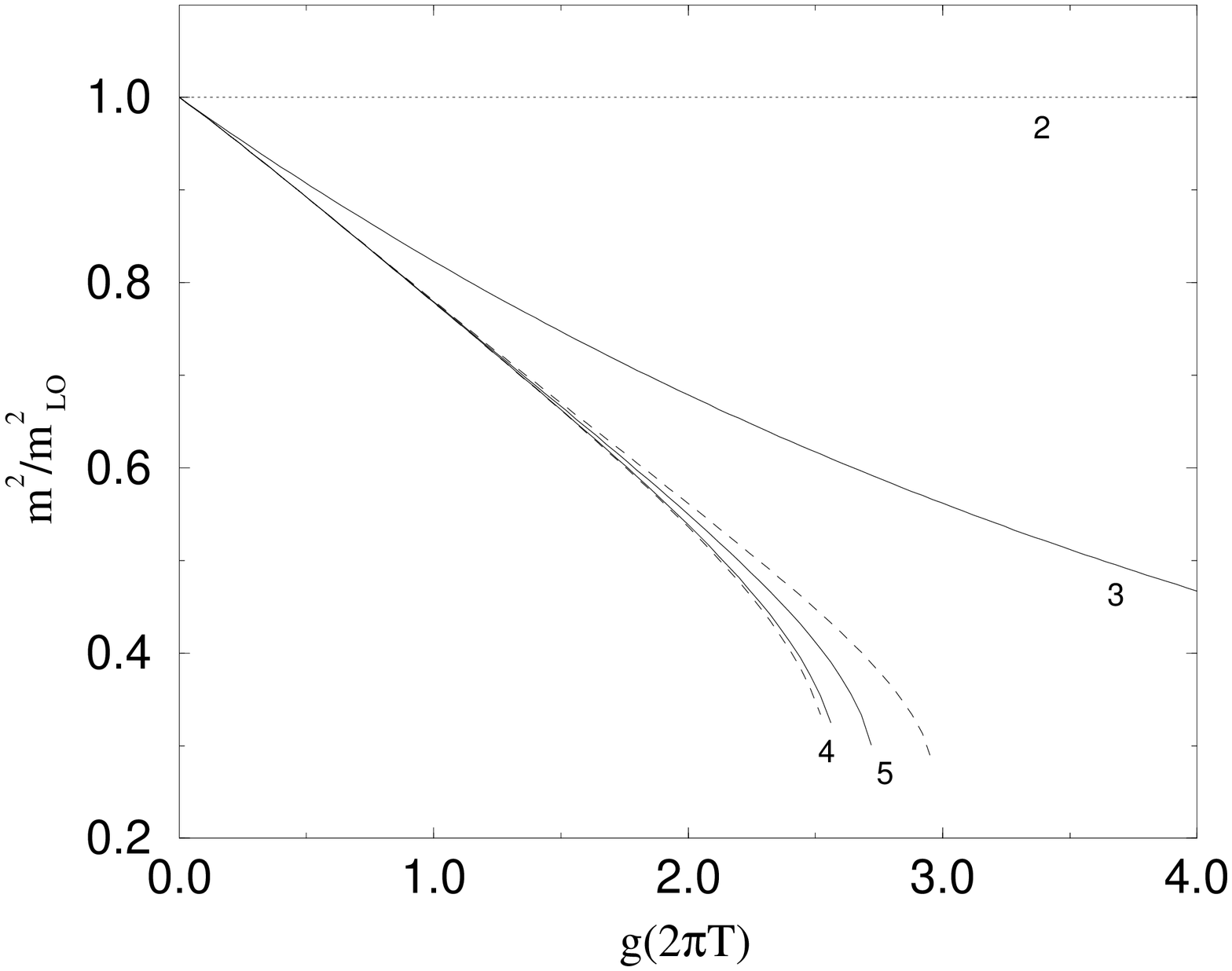,width=7cm,angle=-90}
&
\epsfig{file=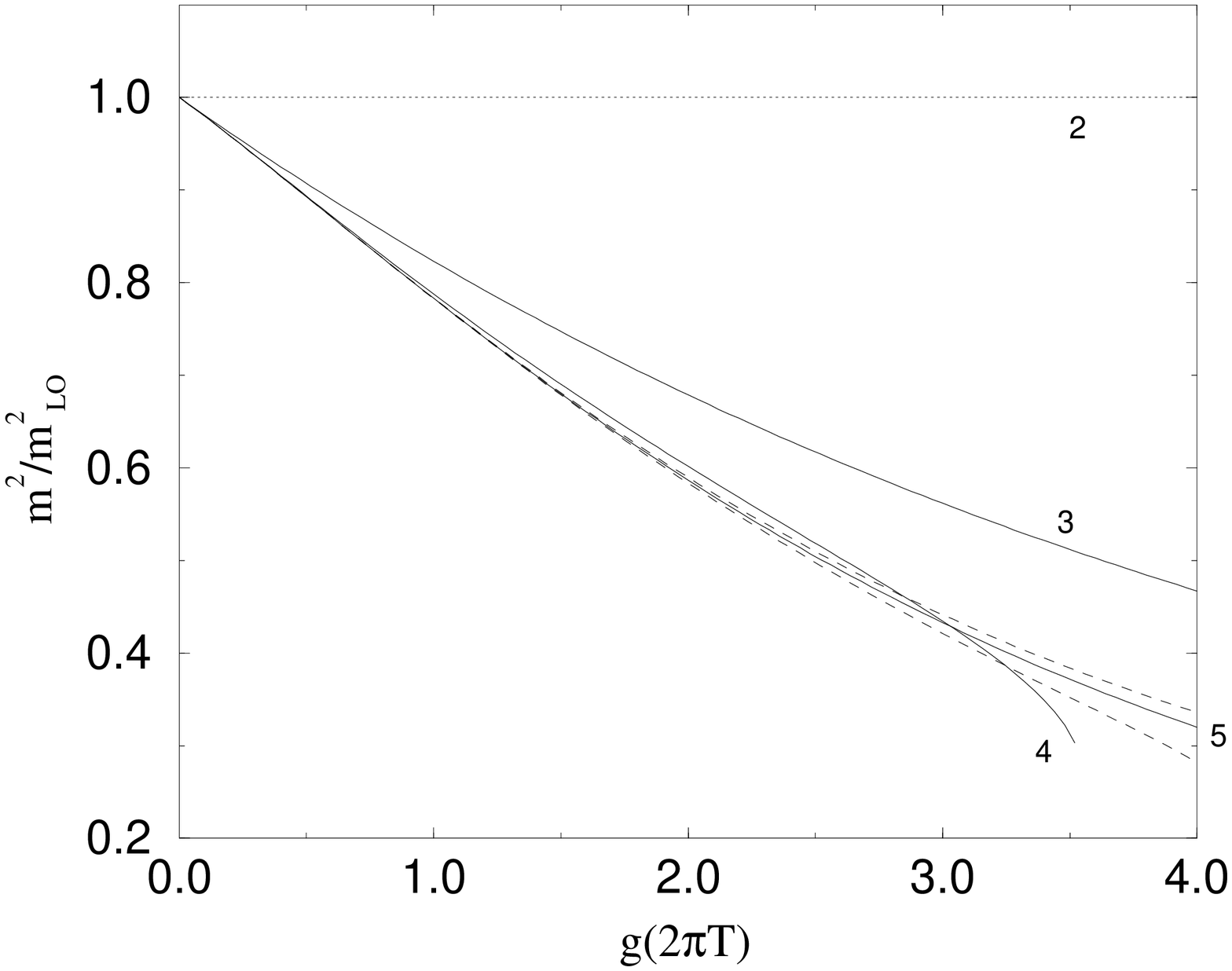,width=7cm,angle=-90}
\vspace*{0.5cm}
\\
\vspace*{0.5cm}
(a) & (b)
\end{tabular}
\caption{$\Phi$-derivable approximation to the screening mass divided by 
the leading order solution $m^2_{\rm LO}=g^2(2\pi T)T^2/24$ as a function of 
$g(2\pi T)$ for (a) $\Lambda=2\pi T$ and (b) $\Lambda=m$. Solid lines are 
solutions to the gap equation truncated at orders $g^n$, $n=2,3,4,5$ with 
$\mu=2\pi T$. Dashed lines are solutions to the gap equation truncated at 
order $g^5$ for $\mu=\pi T$ and $\mu=4\pi T$.}
\end{figure}
The divergent terms of order $g^6$ that were eliminated by the truncation are
\begin{eqnarray}
&& {3\over 2}\left[ {1\over \epsilon^2}+\left( 4L-4\ell+2\right){1\over 
\epsilon}\right]\alpha\hat m^4 \nonumber
\\
&&+\left[ {1\over \epsilon^2}+\left( 6L+2\ell-12\log\hat m+11-24\log 2
-4{\zeta'(-1)\over \zeta(-1)}\right){1\over \epsilon}\right]\alpha^2\hat m^2 
\nonumber
\\
&&-{1\over 12}\left[ {1\over \epsilon^2}+\left( 8L+2\ell-24\log\hat m+
{109\over 5}-48\log 2-20{\zeta'(-1)\over \zeta(-1)}+6{\zeta'(-3)\over 
\zeta(-3)}\right){1\over \epsilon}\right]\alpha^3\,.
\end{eqnarray}
Note  that the double poles in $\epsilon$ do not cancel if we use the leading 
order gap equation $\hat{m}^2=\alpha/6$. Thus they cannot be eliminated by 
adding another term to the thermodynamic potential proportional to 
$(\hat{m}^2-\hat{G})^2$. These ultraviolet divergences arise from 
having truncated the interaction functional $\Phi$ at 3 loops. To eliminate 
them, one would have to include 4-loop terms in $\Phi$. 

In the case of the 2-loop $\Phi$-derivable approximation, we were able to 
eliminate the ultraviolet divergences to all orders by replacing the true 
coupling constant $\alpha$ by a new coupling constant $\bar{\alpha}$ whose 
expansion in $\alpha$ had divergent coefficients but which still had a 
well-defined beta function. If we demand that the resulting beta function has 
the correct one-loop coefficient, the new coupling constant must have the 
form
\begin{equation}
\bar \alpha = \alpha +{A\over \epsilon}\alpha^3+...
\end{equation} 
where $A$ is a constant. By an appropriate choice of this constant, one might 
be able to eliminate one of the terms of order $\bar{g}^6$ with a single 
pole in $\epsilon$. But since we cannot eliminate the double poles, there 
is nothing to be gained by introducing such a coupling constant. 

\subsection {Screening Mass}

The gap equation obtained by varying our finite thermodynamic potential 
(\ref{thpot:fintrunc}) with respect to $m^2$ while holding $\mu$ and 
$\Lambda$ fixed is 
\begin{eqnarray}
\hat{m}^2 &=& \alpha\left[ {1\over 6}-\hat{m}-\left(3L-2\ell+\gamma\right)
\hat{m}^2\right] \nonumber
\\
&&+\alpha^2\left[ -{1\over 3}\left( \ell-2\log \hat{m}+1-4\log 2-
{\zeta'(-1)\over \zeta(-1)}\right) +3\left( \ell+\gamma\right)\hat{m}\right]
\nonumber
\\
&&+\alpha^3\left[ -{1\over 18}{1\over \hat{m}^2}-{1\over 6}\left( \ell-2
\log 2+\gamma\right){1\over \hat{m}}\right] \,.
\label{gap:main}
\end{eqnarray}
The solution to this equation is the 3-loop $\Phi$-derivable approximation 
to the screening mass $m_s$. If we solve the equation iteratively in powers 
of $(\alpha/6)^{1/2}$, it agrees with the weak-coupling expansion 
(\ref{ms-weak}) for the screening mass. The screening mass depends on the 
arbitrary scale $\Lambda$ through the logarithms $\ell$. It also depends on 
the renormalization scale $\mu$ through $\alpha(\mu)$ and $L$. However 
since it agrees with the weak-coupling expansion through order $g^5$, all 
dependence on these arbitrary scales must be of order $g^6$ or higher. 

\begin{figure}[t]
\vspace*{-1cm}
\begin{tabular}{cc}
\epsfig{file=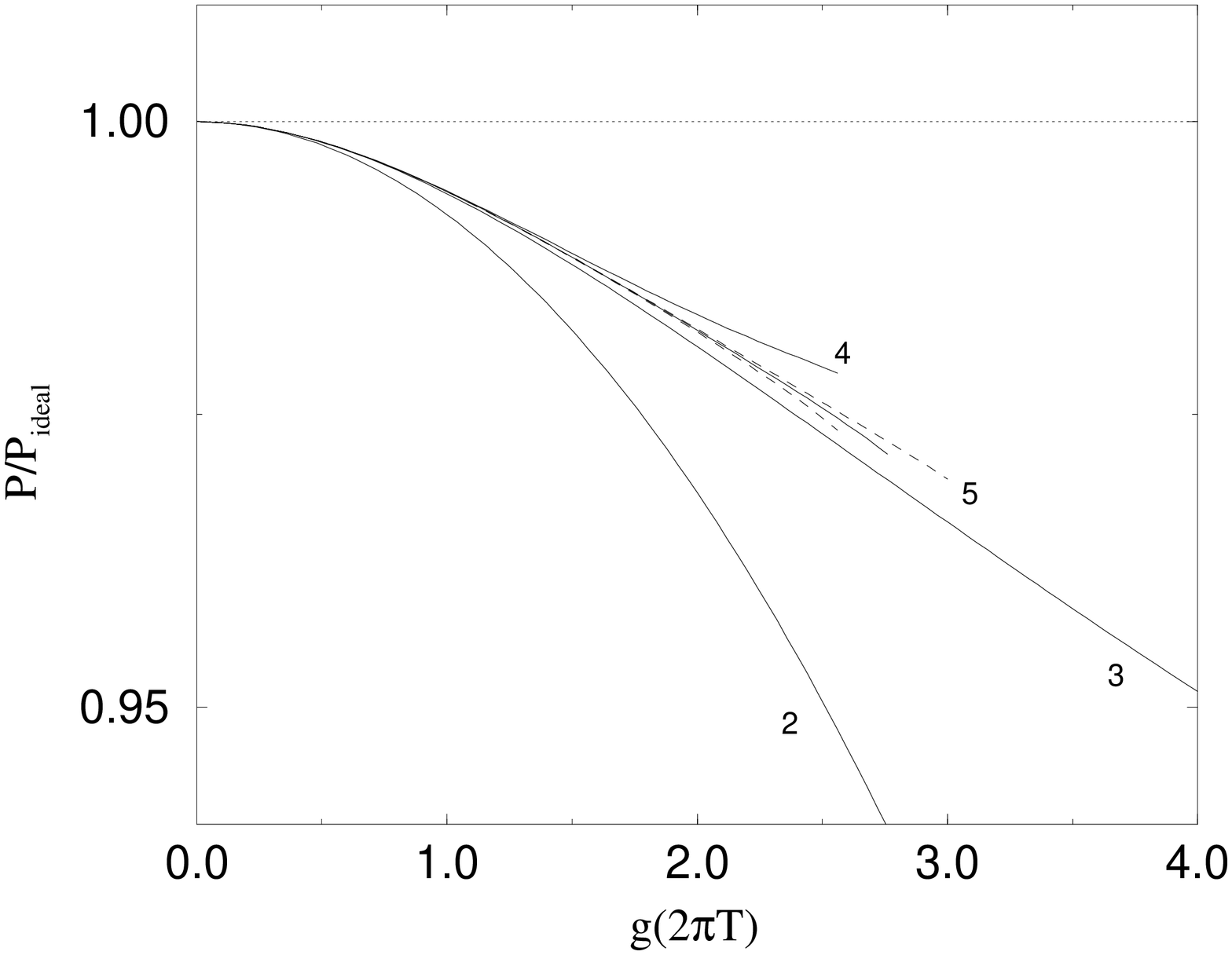,width=7cm,angle=-90}
&
\epsfig{file=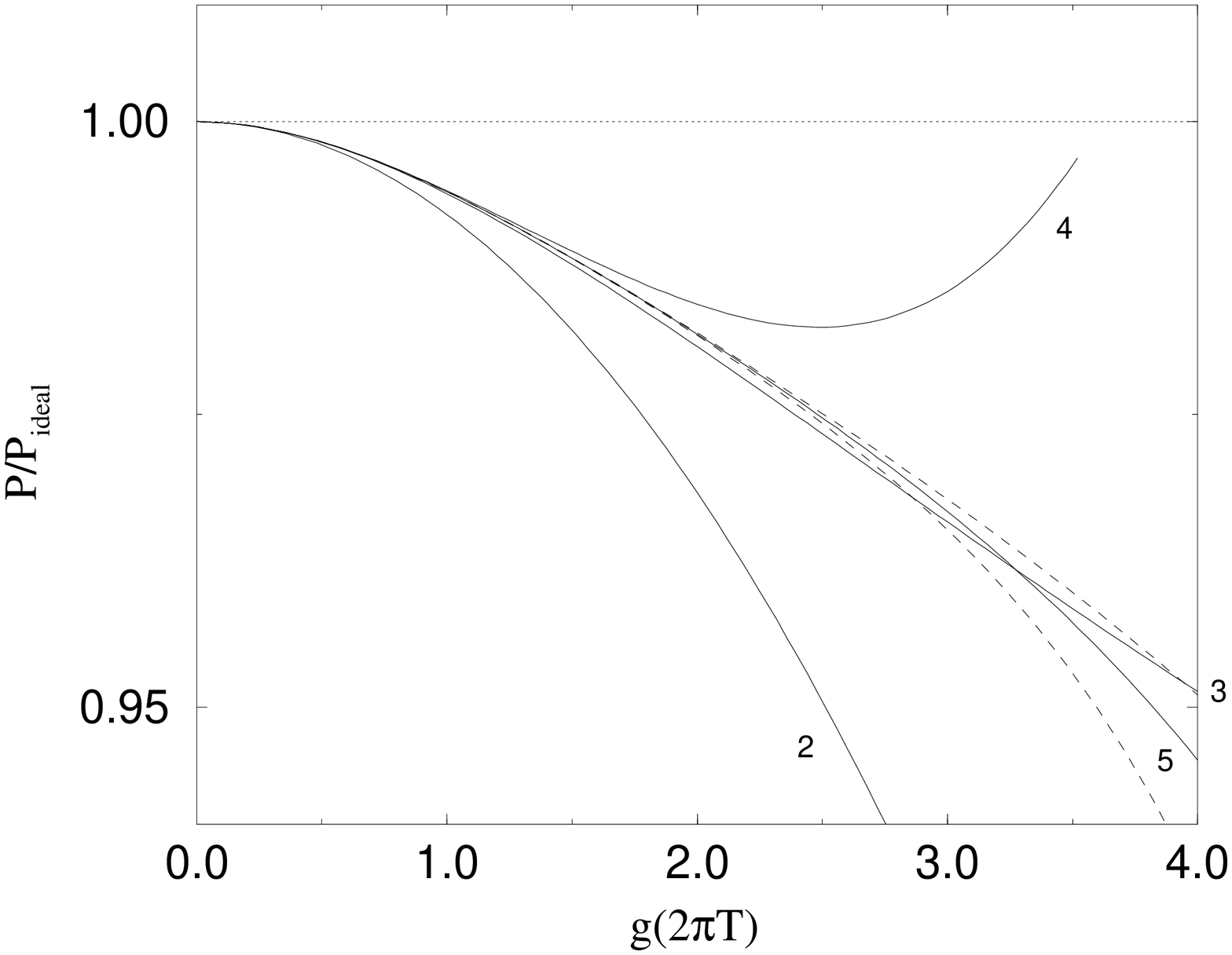,width=7cm,angle=-90}
\vspace*{0.5cm}
\\
\vspace*{0.5cm}
(a) & (b)
\end{tabular}
\caption{$\Phi$-derivable approximation to the pressure divided by that of 
the ideal gas as a function of $g(2\pi T)$ for (a) $\Lambda=2\pi T$ and (b) 
$\Lambda=m$. Solid lines correspond to the truncation of the thermodynamic 
potential at order $g^n$, $n=2,3,4,5$ for $\mu=2\pi T$. Dashed lines 
correspond to the truncation at order $g^5$ for $\mu=\pi T$ and $\mu=4\pi T$.}
\end{figure}
We can obtain a series of successive approximations to the screening mass 
by truncating the gap equation (\ref{gap:main}) after terms of $n^{\rm th}$ 
order in $g$ and $\hat{m}$. In Fig.~4a, we show the solutions for $m^2$ 
divided by the leading order result $m_{\rm LO}^2=g^2(2\pi T)T^2/24$ for 
$\mu=\Lambda=2\pi T$ as a function of $g(2\pi T)$. The solid lines 
correspond to truncations after orders $n=2,3,4,5$. For $n=4$ and $5$, there 
is no solution beyond a critical value of the coupling constant: $g=2.591$ for 
$n=4$ and $g=2.768$ for $n=5$. For larger values of $g$, the thermodynamic 
potential has a runaway minimum $m\to 0$ that comes from the $\alpha^2\log 
\hat{m}$ term in (\ref{thpot:fintrunc}). A similar phenomenum occurs in 
screened perturbation theory at three loops \cite{A-B-S}: if the screening 
mass is used as the variational parameter, the solution to the gap equation 
terminates at nearly the same critical value of $g$. For $g$ below the 
critical value, the screening mass seems to be converging nicely. It is also 
fairly insensitive to the renormalization scale, as shown in Fig.~4a. The 
solution to the gap equation truncated at $5^{\rm th}$ order is shown 
as dashed lines for the two choices $\mu=\pi T$ and $\mu=4\pi T$. 

The simplest way to get a solution for the screening mass that extends beyond 
the critical value of $g$ is to choose the arbitrary scale $\Lambda$ to be 
proportional to $m$. This changes the gap equation to
\begin{eqnarray}
\hat{m}^2 &=& \alpha\left[ {1\over 6}-\hat{m}-\left(3L-2\ell-{1\over 2}
+\gamma\right)\hat{m}^2\right] \nonumber
\\
&&+\alpha^2\left[ -{1\over 3}\left( \ell-2\log \hat{m}+{3\over 2}-4\log 2-
{\zeta'(-1)\over \zeta(-1)}\right) +3\left( \ell+\gamma+{1\over 3}
\right)\hat{m}\right]
\nonumber
\\
&&+\alpha^3\left[ -{1\over 24}{1\over \hat{m}^2}-{1\over 6}\left( \ell+1-2
\log 2+\gamma\right){1\over \hat{m}}\right] \,.
\label{gap:Lambda=m}
\end{eqnarray}
When truncated at $2^{\rm nd}$ and $3^{\rm rd}$ order in $g$ and $\hat{m}$, 
the gap equations (\ref{gap:main}) and (\ref{gap:Lambda=m}) are identical. 
The solution to the gap equation (\ref{gap:Lambda=m}) truncated at 
$n^{\rm th}$ order and with $\mu=2\pi T$ and $\Lambda=m$ are shown as solid 
lines in Fig.~4b. The solution for $n=4$ terminates at $g=3.551$ for 
$\mu=2\pi T$. The solution for $n=5$ continues well beyond $g(2\pi T)=4$. 
The dashed lines are the $5^{\rm th}$ order truncation with $\Lambda=m$ for 
two choices of the renormalization scale: $\mu=\pi T$ and $\mu=4\pi T$. 
\begin{figure}[t]
\vspace*{-1cm}
\begin{tabular}{cc}
\epsfig{file=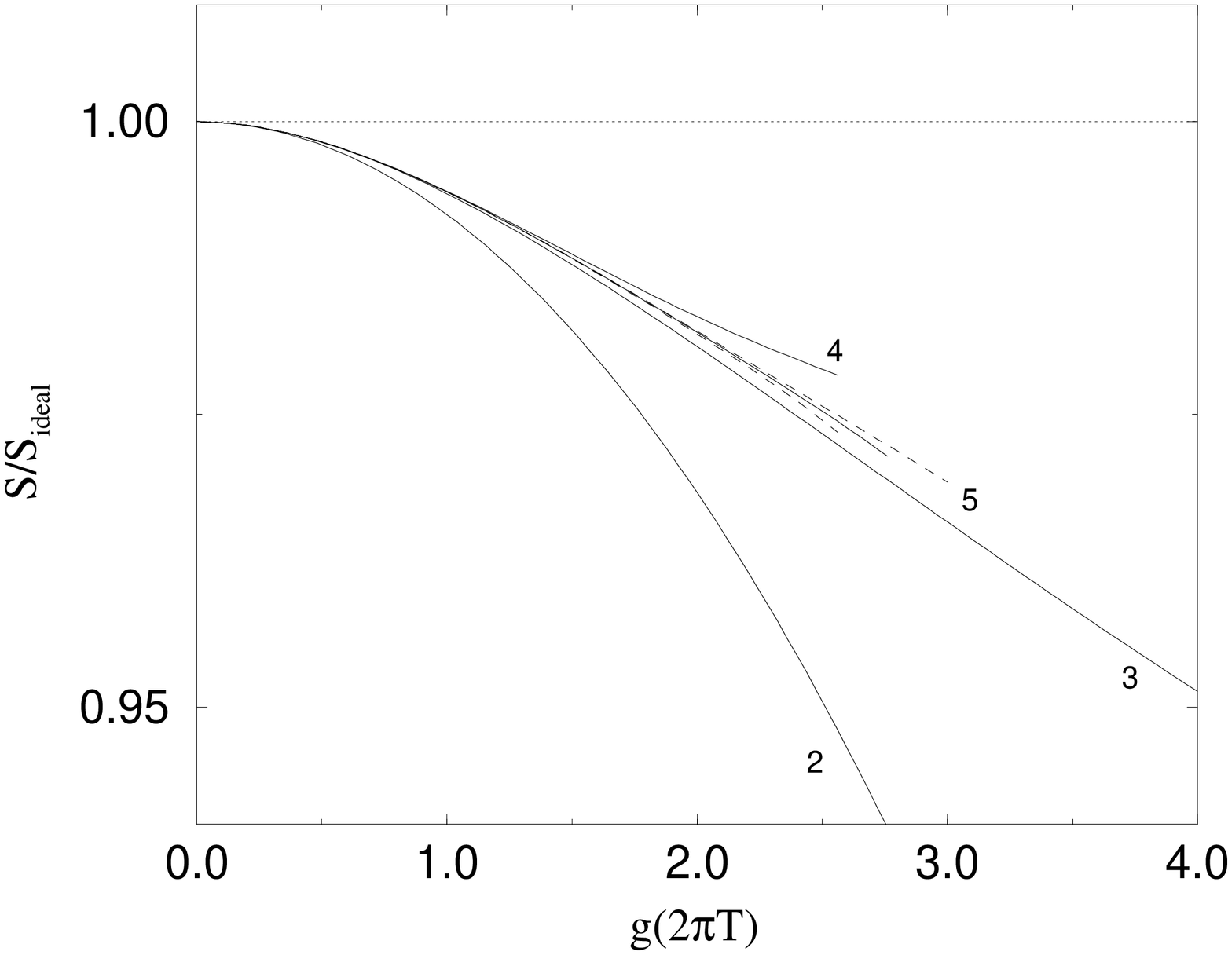,width=7cm,angle=-90}
&
\epsfig{file=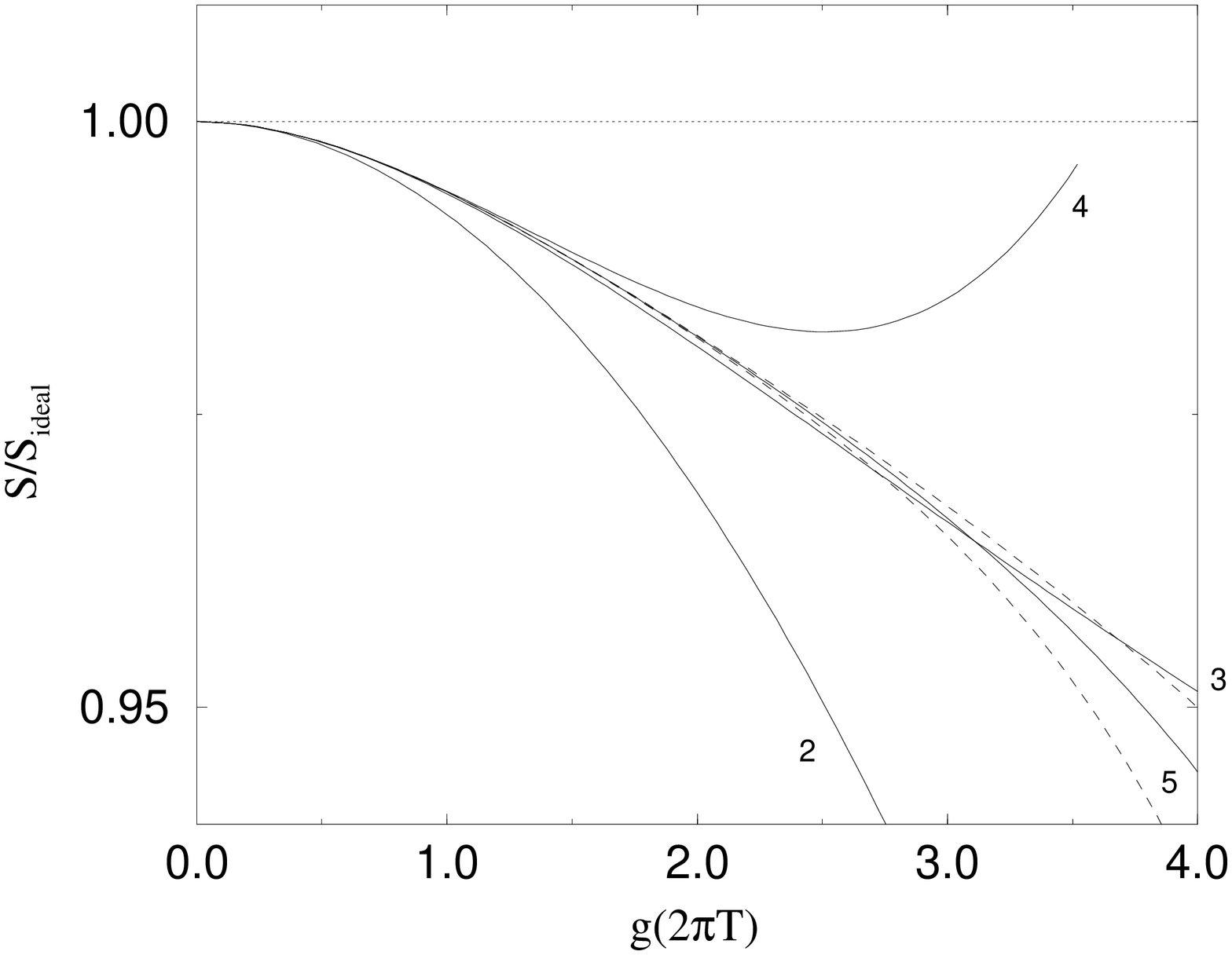,width=7cm,angle=-90}
\vspace*{0.5cm}
\\
\vspace*{0.5cm}
(a) & (b)
\end{tabular}
\caption{$\Phi$-derivable approximation to the entropy divided by that of 
the ideal gas as a function of $g(2\pi T)$ for (a) $\Lambda=2\pi T$ and (b) 
$\Lambda=m$. Solid lines correspond to the truncation of the entropy at 
order $g^n$, $n=2,3,4,5$ for $\mu=2\pi T$. Dashed lines correspond to the 
truncation at order $g^5$ for $\mu=\pi T$ and $\mu=4\pi T$.}
\end{figure}

\subsection {Thermodynamic Functions}

The pressure, given by (\ref{pressure}), is obtained by inserting the 
solution to the gap equation (\ref{gap:main}) into the thermodynamic 
potential (\ref{thpot:fintrunc}). The pressure truncated at $n^{\rm th}$ 
order in $g$ and $\hat{m}$ is obtained by using the $n^{\rm th}$ order 
truncations of (\ref{gap:main}) and (\ref{thpot:fintrunc}). In Fig.~5a, we 
show the $n^{\rm th}$ order truncations of the pressure divided by that of 
the ideal gas for $\Lambda=2\pi T$ and $\mu=2\pi T$. We show also the change 
in the $5^{\rm th}$ order truncation from varying the renormalization scale 
$\mu$. Since the solutions to the $4^{\rm th}$ order and $5^{\rm th}$ order 
gap equations cannot be continued beyond the critical values 
$g(2\pi T)=2.591$ and $2.768$ respectively, the corresponding lines for 
the pressure cannot be extended past these values. The variations of the 
pressure are much smaller than one might expect from the variations of the 
screening mass in Fig.~4a. We see an obvious improvement in the convergence 
compared to the weak-coupling expansion. At $5^{\rm th}$ order, and up to 
the critical values of $g(2\pi T)$ starting around $2.5$, the variations 
from the renormalization scale are much smaller than those of the 
weak-coupling expansion. Fig.~5b shows the pressure truncated at the same 
orders but with the scale $\Lambda=m$. The pressure is obtained by inserting 
the solution to the gap equation (\ref{gap:Lambda=m}) into the thermodynamic 
potential (\ref{thpot:fintrunc}). The line for $n=5$ continues well beyond 
$g(2\pi T)=4$, but for values of $g(2\pi T)$ around $2.5$ the convergence 
between $n=3,4$ and $5$ is not as good as in the previous case of 
$\Lambda=2\pi T$. Up to $g(2\pi T)=4$, the changes in the $5^{\rm th}$ 
order truncation from varying the renormalization scale $\mu$ appears to be 
much smaller than in the weak-coupling expansion. 

The diagrammatic entropy is given by Eq.~(\ref{entropdens}), with $\mu$ and 
$\alpha(\mu)$ held fixed. We have to consider two cases giving 
different expressions of the entropy, the case with $\Lambda=c(2\pi T)$ 
and the case with $\Lambda=c\; m$. In the first case, upon differentiating 
the thermodynamic potential (\ref{thpot:fintrunc}) with respect to $T$, the
$5^{\rm th}$ order expression for the entropy is  
\begin{eqnarray}
{1\over 15}\left( {{\cal S}\over {\cal S}_{\rm ideal}}-1\right) &=& 
{1\over 4}\left[ -2\hat{m}^2+4\hat{m}^3-9\hat{m}^4\right] \nonumber
\\
&&+\alpha\left[ \left( \ell-2\log \hat{m}+3-4\log 2-{\zeta'(-1)
\over \zeta(-1)}\right)\hat{m}^2-3\left( \ell+\gamma\right) 
\hat{m}^3\right] \nonumber
\\
&&+\alpha^2\left[ -{1\over 6}\left( \ell-4\log \hat{m}+{119\over 30}
-8\log 2-4{\zeta'(-1)\over \zeta(-1)}+{\zeta'(-3)\over \zeta(-3)}\right)
\right. \nonumber
\\
&&\hspace{1cm}
\left.+{3\over 2}\left( \ell-2\log 2+\gamma\right)\hat{m}
\right] \,,
\label{entrop}
\end{eqnarray}
where ${\cal S}_{\rm ideal}=(2\pi^2/45)T^3$, and $\hat{m}$ is the solution of 
the gap equation (\ref{gap:main}). Fig.~6a displays the good convergence of 
the $n^{\rm th}$ order truncation of the entropy for $\mu=2\pi T$ and 
$\Lambda=2\pi T$, and the small variations from the renormalization scale 
$\mu$. The $5^{\rm th}$ order expression for the entropy for the choice 
$\Lambda=c\; m$ is given by 
\begin{eqnarray}
{1\over 15}\left( {{\cal S}\over {\cal S}_{\rm ideal}}-1\right) &=& 
{1\over 4}\left[ -2\hat{m}^2+4\hat{m}^3-3\hat{m}^4\right] \nonumber
\\
&&+\alpha\left[ \left( \ell-2\log \hat{m}+{5\over 2}-4\log 2-{\zeta'(-1)
\over \zeta(-1)}\right)\hat{m}^2-3\left( \ell-1+\gamma\right) 
\hat{m}^3\right] \nonumber
\\
&&+\alpha^2\left[ -{1\over 6}\left( \ell-4\log \hat{m}+{223\over 60}
-8\log 2-4{\zeta'(-1)\over \zeta(-1)}+{\zeta'(-3)\over \zeta(-3)}\right)
\right. \nonumber
\\
&&\hspace{1cm}
\left.+{3\over 2}\left( \ell-{1\over 3}-2\log 2+\gamma\right)\hat{m}
\right] \,,
\label{entropm}
\end{eqnarray}
with $\hat{m}$ the solution to the gap equation (\ref{gap:Lambda=m}) 
and $\ell=\log(m/2\pi T)$. Fig.~6b shows the $n^{\rm th}$ order truncations 
of the entropy for $\Lambda=m$. We conclude again, as in the case 
of the pressure, that around the critical values of $g(2\pi T)$ the 
convergence is not as good as in the case of $\Lambda=2\pi T$. 
However the changes of the $5^{\rm th}$ order truncation with respect to 
the variations of the renormalization scale $\mu$ are still relatively small.

\section{Conclusions}

We have developed a systematic method for solving $\Phi$-derivable 
approximations for the massless $\phi^4$ field theory. The method involves 
expanding sum-integrals systematically in powers of $m/T$, where $m$ is the 
screening mass. When expanded in powers of $m/T$, the variational equations 
for the self-energy $\Pi(P)$ have a recursive structure that allows then to 
be solved in terms of the variable $m$. This reduces the problem to one that 
involves the single variational parameter $m$. Upon truncating the expansion 
in $m/T$, we obtain a gap equation for $m$ that could easily be solved 
numerically were it not for the presence of ultraviolet divergences. Some 
of the ultraviolet divergences vanish at the variational point, and can be 
eliminated by a redefinition of the thermodynamic potential. However there 
are other ultraviolet divergences that do not vanish at the variational point. 
They are generated by the truncation error in defining the interaction 
functional $\Phi[\Pi]$. They limit the accuracy to which the $\Phi$-derivable 
approximation can be solved. 

In the case of the 2-loop $\Phi$-derivable approximation, the unavoidable 
ultraviolet divergences appear at order $g^4$. They can be eliminated to all 
orders by a trick that involves introducing a new coupling constant $\bar{g}$ 
that runs with the wrong beta function. However this trick does not generalize 
to higher orders in the loop expansion. If one expresses the thermodynamic 
potential in terms of the true coupling constant $g$, the best one can do in 
constructing a finite thermodynamic potential is to truncate after terms of 
third order in $m/T$. 

In the 3-loop $\Phi$-derivable approximation, the unavoidable ultraviolet 
divergences appear at order $g^6$. We constructed a finite thermodynamic 
potential by adding terms proportional to the square of the gap equation to 
eliminate the ultraviolet divergences that vanish at the variational point 
and then truncating after terms of $5^{\rm th}$ order in $g$ and $m/T$. 
Truncations of the finite thermodynamic potential at $n^{\rm th}$ order in 
$g$ and $m/T$, where $n=2,3,4,5$, define a sequence of successive variational 
approximations to the thermodynamics. The truncation after terms of 
$3^{\rm rd}$ order in $g$ and $m/T$ reproduces the finite thermodynamic 
potential defined by the 2-loop $\Phi$-derivable approximation. Our solution 
of the 3-loop $\Phi$-derivable approximation allows the $4^{\rm th}$ order 
and $5^{\rm th}$ order approximations to be constructed. The $6^{\rm th}$ 
order and $7^{\rm th}$ order approximations could be constructed by applying 
our method to the 4-loop $\Phi$-derivable approximation. The only obstacle 
is evaluating the scalar sum-integrals that would arise. The most difficult 
of the sum-integrals are the same ones  that would need to be evaluated to 
extend the weak-coupling expansion to order $g^6$.

We calculated numerically the thermodynamic functions defined by the finite 
thermodynamic potential truncated at $n^{\rm th}$ order in $g$ and $m/T$ for 
$n=2,3,4,$ and $5$. One problem is that in the cases $n=4$ and $5$, the 
solution to the gap equation for $m$ cannot be extended beyond a critical 
value of $g$ that is near $g=2.5$ if we choose the renormalization scale 
$\mu=2\pi T$. For values of $g$ below the critical value, the successive 
approximations seem to be converging and are rather insensitive to 
variations in the renormalization scale. 

Our construction of a finite thermodynamic potential involved a prescription 
for removing from the bare thermodynamic potential $\Omega_0$ ultraviolet 
divergences that vanish at the variational point. Our prescription was to 
add the two terms in (\ref{addterm}) and (\ref{addterm2}) that are both 
proportional to the square of the gap equation. We had no compelling 
motivation for this prescription beyond its success in the 2-loop 
$\Phi$-derivable approximation. This prescription introduces an ambiguity 
into the finite thermodynamic potential (\ref{thpot:fintrunc}). We can 
feel free to add additional terms proportional to $(\hat{m}^2-\hat{G})^2$, 
where $\hat{G}$ is the right side of the finite gap equation given in 
(\ref{gap:main}) or (\ref{gap:Lambda=m}). If we avoid large coefficients for 
these additional terms, they do not have a large effect numerically for 
$g$ less than about $2.5$. However the ambiguity emphasizes that our 
prescription was rather ad hoc. It would be desirable to have a much more 
compelling prescription for eliminating the ultraviolet divergences that 
vanish at the variational point. 

Our strategy for solving $\Phi$-derivable approximations should also be 
applicable to more complicated field theories, such as gauge theories. In the 
case of $QCD$, Blaizot, Iancu and Rebhan \cite{B-I-R} have solved the 2-loop 
$\Phi$-derivable equations for the self-energy tensor $\Pi^{\mu\nu}$ to 
leading order in $g_s$ in terms of a Debye screening mass parameter $m_D$. 
They did not determine $m_D$ variationally, but simply used the asymptotic 
weak-coupling expression $m_D=g_sT$. The resulting approximation reproduces 
the thermodynamic functions through order $g_s^2$. They called it a 
``leading order'' $({\rm LO})$ $HTL$ approximation. The application of our 
method to gauge theories would involve expanding sum-integrals systematically 
in powers of $g_s$ and $m_D/T$. If applied to the 2-loop $\Phi$-derivable 
approximation for $QCD$, it should reproduce the leading order solution for 
$\Pi^{\mu\nu}(P)$ in Ref.~\cite{B-I-R}, but it should give a gap 
equation for $m_D$ that allows it to be determined variationally. By 
truncating after terms of $3^{\rm rd}$ order in $g_s$ and $m_D/T$, one should 
be able to improve the accuracy of the thermodynamic functions to order 
$g_s^3$. At higher orders in $g_s$ and $m_D/T$, the 2-loop 
$\Phi$-derivable approximation will be plagued by ultraviolet divergences 
associated with the truncation of $\Phi$ as well as by gauge dependence 
associated with the use of the gauge-dependent gluon propagator as a 
variational function. To improve the accuracy while avoiding these problems, 
one would have to apply our method to the 3-loop $\Phi$-derivable 
approximation to $QCD$. This would allow the accuracy to be improved to 
orders $g_s^4$ and $g_s^5$. Any further improvement in the accuracy would 
be very difficult. One problem is that there are new sum-integrals that 
enter at order $g_s^6$ that have never been evaluated. A more serious problem 
is that the 3-loop free energy is sensitive to the nonperturbative momentum 
scale $g_s^2T$ associated with magnetic screening. This will probably cause 
our method for reducing the $\Phi$-derivable approximation to a 
single-parameter variational problem to breakdown.

Blaizot, Iancu, and Rebhan \cite{B-I-R} have introduced an improvement on 
their ${\rm LO}$ $HTL$ approximation that reproduces the thermodynamic 
functions to order $g_s^3$. They call it a ``next-to-leading order'' 
$({\rm NLO})$ $HTL$ approximation. It reduces essentially to the substitution
\begin{equation}
m_D^2\longrightarrow {g_s^2T^2\over 1+3g_s/\pi}
\label{substit}
\end{equation} 
in their ${\rm LO}$ $HTL$ approximation. The simpler prescription 
$m_D^2\to g_s^2T^2(1-3g_s/\pi)$ gives the same parametric improvement in the 
accuracy, but is a disaster phenomenologically, giving a negative screening 
mass unless $\alpha_s<0.087$. The ad hoc prescription (\ref{substit}) gives 
phenomenologically acceptable results that are not very different from the 
${\rm LO}$ $HTL$ approximation of Ref.~\cite{B-I-R}. We argue that such a 
prescription is unlikely to arise in any systematically improvable 
approximation to the thermodynamics. Our solution to the $\Phi$-derivable 
approximation to the massless $\phi^4$ field theory suggests an alternative 
prescription. The truncation of a gap equation after terms that are 
$3^{\rm rd}$ order in $g_s$ and $m_D/T$ as in (\ref{simplegap}) would give  
\begin{equation}
m_D\longrightarrow g_sT\left[ \left( 1+{9\over 4\pi^2}g_s^2\right)^{1/2}
-{3\over 2\pi}g_s\right]\,.
\end{equation}   

In summary, we have developed a new method for solving the 3-loop 
$\Phi$-derivable approximation for the massless $\phi^4$ field theory. This 
method can be used to construct a sequence of systematically improvable 
approximations to the thermodynamic functions. This sequence of 
approximations seems to be stable and converging for $g(2\pi T)$ less than 
$2.5$, and also insensitive to the choice of renormalization scale. If this 
method can be adapted to gauge theories, it may provide a practical solution 
to the longstanding problem of large radiative corrections to the 
thermodynamic functions.

\section*{Acknowledgments}
This work was supported in part by the U.~S. Department of
Energy under grant DE-FG02-91-ER40690. We thank J.~O.~Andersen and 
M.~Strickland for valuable discussions.

\appendix\bigskip\renewcommand{\theequation}{\thesection.\arabic{equation}}

\newpage

\section{Expansions in $m/T$}
\setcounter{equation}{0}
\label{app:movert}

In this Appendix, we carry out the expansions of sum-integrals in $m/T$. We 
first illustrate the method by applying it to one-loop sum-integrals for a 
particle of mass $m$. We then apply it to sum-integrals with a momentum 
dependent self-energy $\Pi(P)$.

\subsubsection{Simple examples}

To illustrate the $m/T$ expansion in the simplest context, we carry it out 
for the one-loop sum-integrals that appear in the 2-loop $\Phi$-derivable 
approximation. The sum-integrals over $P=(2\pi nT, {\bf p})$ in 
(\ref{thpot2:1}) and (\ref{gap2:1}) include sums over all Matsubara 
frequencies $n$ and integrals over all momenta ${\bf p}$. There are two 
important momentum scales: the {\it hard} scale $2 \pi T$ and the {\it soft} 
scale $m$. The hard region for the momentum $P= (2 \pi nT, {\bf p})$ includes 
$n \neq 0$ for all ${\bf p}$ and also $n = 0$ with ${\bf p}$ of order $T$. 
The soft region is $n = 0$ and ${\bf p}$ of order $m$.  
We will evaluate the contributions to the sum-integrals from these 
two momentum regions separately.

The tadpole sum-integral in (\ref{gap2:1}) can be
separated into contributions from hard and soft regions, which we denote
by the superscript $(h)$ and $(s)$:
\begin{equation}
\sumint_P {1 \over P^2 + m^2} = 
\sumint^{(h)}_P {1 \over P^2 + m^2} 
+ \sumint^{(s)}_P {1 \over P^2 + m^2}\,.
\label{tadpole2:1}
\end{equation}
In the hard region, $P^2$ is of order $T^2$, 
so we can expand in powers of $m^2/P^2$:
\begin{eqnarray}
\sumint^{(h)}_P {1 \over P^2 + m^2} &=& 
\sumint_P \left( {1 \over P^2} - {m^2 \over (P^2)^2} 
	+ \sum_{n=2}^\infty (-1)^n {m^{2n}  \over (P^2)^{n+1}} \right).
\label{tadpole2:h}
\end{eqnarray}
The contribution from the soft region can be written
\begin{equation}
\sumint^{(s)}_P {1 \over P^2 + m^2} = T \int_{\bf p} {1 \over p^2 + m^2}\,.
\label{tadpole2:s}
\end{equation}
One could accomplish the separation of the hard and soft momentum 
regions by introducing a separation scale $\Lambda$
satisfying $m \ll \Lambda \ll 2 \pi T$.  The dependence on $\Lambda$ 
would then cancel between (\ref{tadpole2:h}) and (\ref{tadpole2:s}). 
However, if we use dimensional regularization, we can get the correct 
answer more easily by taking the limit $\Lambda \to 0$ in 
(\ref{tadpole2:h}) and $\Lambda \to \infty$ in (\ref{tadpole2:s}).
The complete expansion of the sum-integral in powers of $m/T$ 
is then just the sum of (\ref{tadpole2:h}) and (\ref{tadpole2:s}):
\begin{eqnarray}
\sumint_P {1 \over P^2 + m^2} &=& 
\sumint_P {1 \over P^2} + T I_1 - m^2 \sumint_P {1 \over (P^2)^2}
+ \sum^\infty_{n=2} (-1)^n m^{2n} \sumint_P {1 \over (P^2)^{n+1}} \,,
\label{tadpole2:2}
\end{eqnarray}
where $I_1$ is the momentum integral on the right side of
(\ref{tadpole2:s}).

The logarithmic sum-integral in (\ref{thpot2:1}) can also be
separated into contributions from hard and soft regions,
as in (\ref{tadpole2:1}).  In the hard contribution, 
we can expand in powers of $m^2/P^2$: 
\begin{eqnarray}
\sumint^{(h)}_P \log (P^2 + m^2) & = &
\sumint_P \left( \log P^2 + {m^2 \over P^2} 
	- {1 \over 2} {m^4 \over (P^2)^2}
	+ \sum^\infty_{n = 3} {(-1)^{n+1} \over n} {m^{2n} \over (P^2)^n} 
	\right) \,.
\label{logint:h}
\end{eqnarray}
The soft contribution is 
\begin{eqnarray}
\sumint^{(s)}_P \log (P^2 + m^2) & = & T \int_{\bf p} \log (p^2 +
m^2)  \,.
\label{logint:s}
\end{eqnarray}
The complete expansion of the sum-integral in powers of $m/T$ is
the sum of (\ref{logint:h}) and (\ref{logint:s}):
\begin{eqnarray}
\sumint_P \log (P^2 + m^2) & = &
\sumint_P \log P^2 + m^2 \sumint_P {1 \over P^2} + T(-I_0')
\nonumber 
\\
&& - {1 \over 2} m^4 \sumint_P {1 \over (P^2)^2}
+ \sum^\infty_{n = 3}{(-1)^{n+1} \over n} ~ m^{2n}
        \sumint_P {1 \over (P^2)^n}  \,,
\label{logint:2}
\end{eqnarray}
where $-I_0'$ is the momentum integral on the right side of
(\ref{logint:s}).

\subsubsection{Tadpole sum-integral}

The simplest sum-integral with a momentum-dependent self-energy is the 
tadpole sum-integral, which appears in the expression (\ref{screen:2})
for the screening mass.  The tadpole sum-integral can be
separated into contributions from hard and soft regions, which we denote
by a superscript $(h)$ or $(s)$:
\begin{equation}
\sumint_Q {1 \over Q^2 + \Pi (Q)} = 
\sumint^{(h)}_Q {1 \over Q^2 + \Pi(Q)} 
+ \sumint^{(s)}_Q {1 \over Q^2 + \Pi (Q)} \,.
\end{equation}
In the hard region, $Q^2$ is of order $T^2$ while $\Pi$ is of order
$m^2$, so we can expand in powers of $\Pi(Q^2)$.  Inserting the
expansion (\ref{Pi-hard}), we obtain
\begin{eqnarray}
\sumint^{(h)}_Q {1 \over Q^2 + \Pi (Q)} &=& 
\sumint_Q \left( {1 \over Q^2} - {m^2 \over (Q^2)^2}
		+ {m^4 \over (Q^2)^3}  \right)
\nonumber
\\
&& - (g_0 \mu^{-\epsilon})^4 
	\sumint_Q {\Pi_{4,0}(Q) + \Pi_{4,1}(Q) \over (Q^2)^2}
+ {\cal O} (g^6) \,.
\label{tadpole3:h}
\end{eqnarray}
The first term on the right side is proportional to $T^{2-2\epsilon}$. 
Taking $m$ to be of order $g$, the others are suppressed by $g^2$, $g^4$, 
$g^4$, and $g^5$, respectively, in the limit $\epsilon\to 0$. The 
contribution from the soft region can be written
\begin{equation}
\sumint_Q^{(s)} {1 \over Q^2 + \Pi (Q)} 
= T \int_{\bf q} {1 \over q^2 + m^2 + \sigma(q)}\,.
\end{equation}
In this region, $q^2$ is of order $m^2$, which is the first term in the
expansion (\ref{Pi-soft}) of $\Pi (0, p)$.  Expanding in powers of the
higher terms, we have
\begin{eqnarray}
\sumint_Q^{(s)} {1 \over Q^2 + \Pi (Q)} &=& 
T I_1
- (g_0 \mu^{-\epsilon})^4 T 
	\int_{\bf q} {\sigma_{4,-2}(q) + \sigma_{4,0}(q) \over (q^2 +
m^2)^2}
\nonumber
\\
&& 
- (g_0 \mu^{-\epsilon})^8 T \int_{\bf q} 
\left( {\sigma_{8,-4}(q)\over (q^2 + m^2)^2}-
{\sigma_{4,-2}^2(q) \over (q^2 + m^2)^3} \right) + {\cal O}(g^6) \,.
\label{tadpole3:s}
\end{eqnarray}
The first term on the right side is proportional to $m^{1-2\epsilon}T$. In 
the limit $\epsilon\to 0$, it is of order $g$ when $m$ is of order $gT$. 
In this limit the other terms are of order $g^3$, $g^5$, $g^5$, and $g^5$, 
respectively. The error estimate ${\cal O}(g^6)$ on the sum-integrals 
(\ref{tadpole3:h}) and (\ref{tadpole3:s}) refers to their orders in $g$ in 
the limit $\epsilon\to 0$. 

The complete expression for the tadpole sum-integral is the sum of the
hard and soft contributions in (\ref{tadpole3:h}) and (\ref{tadpole3:s}). 
The integrals and sum-integrals that do not involve the self energy
functions $\Pi_{n,m}(Q)$ and $\sigma_{n,m}(q)$ are given 
in Appendices~~\ref{app:sumint} and \ref{app:int}.

\subsubsection {Sunset sum-integral for hard $P$}

The sunset sum-integral (\ref{ISun}) depends on the external momentum $P$, 
and the expansion in powers of $m/T$ is different for hard $P$ and soft $P$. 
We first consider the case of hard $P$.  There are 3 different sum-integration
regions that must be considered.  The momenta $Q$ and $R$ can both be
hard, one can be hard and the other soft, or they both can be soft. We
denote these regions by superscripts $(hh)$, $(hs)$, and $(ss)$, respectively.

In the $(hh)$ region, we can expand in powers of the self-energy.  Only
the $m^2$ term in the self-energy (\ref{Pi-hard}) contributes up to order
$g^4 T^2$:
\begin{equation}
\Im _{\rm sun}^{(hh)} (P) = 
\sumint_{QR} \left( {1 \over Q^2 R^2 S^2} 
	- {3 m^2 \over (Q^2)^2 R^2 S^2} \right)
+ {\cal O}(g^4) \,.
\label{ISunh:hh}
\end{equation}

We next consider the $(hs)$ region for hard $P$.  Taking into account that
any of the three momenta $Q$, $R$, and $S= -(P+Q+R)$ can be soft, the
sum-integral can be written
\begin{eqnarray}
\Im _{\rm sun}^{(hs)} (P) &=& 
3 T \int_{\bf r} \sumint_{Q} {1 \over Q^2 + \Pi(Q)} 
	{1 \over r^2 + m^2 + \sigma (r)}
	{1 \over (P + Q)^2 + 2({\bf p} + {\bf q}) \!\cdot\! {\bf r} 
		+ r^2 + \Pi (S)} \,,
\end{eqnarray}
where $S = -(P+Q)-(0,{\bf r})$.  The $m / T$ expansion is obtained by
expanding the sum-integrand around $1/ [Q^2 (r^2 + m^2) (P + Q )^2]$.
The first few terms are
\begin{eqnarray}
\Im _{\rm sun}^{(hs)} (P) &=& 
3 T I_1 \sumint_Q {1 \over Q^2 (P + Q)^2} 
- 3 T \left(I_0+m^2I_1\right)\sumint_Q {1 \over (Q^2)^2 (P + Q)^2} \nonumber
\\
&& + {12\over d}T\left(I_0-m^2I_1\right)\sumint_Q{{\bf q}^2 \over (Q^2)^3 
(P + Q)^2}\nonumber
\\
&& - 3 (g_0 \mu^{-\epsilon})^4 T 
	\int_{\bf r} {\sigma_{4,-2}(r) \over (r^2 + m^2)^2} 
	\sumint_Q {1 \over Q^2 (P+Q)^2}  + {\cal O} (g^4) \,.
\label{ISunh:hs}
\end{eqnarray}
With dimensional regularization, the integral $I_0$ vanishes. The factor of 
$1/d$ multiplying one of the sum-integrals comes from averaging over the 
angles of ${\bf r}$ in $d$ dimensions:
$[ ({\bf p} + {\bf q}) \cdot {\bf r}]^2 
	\to ({\bf p} + {\bf q})^2 {\bf r}^2/d$. 
The change of variables $Q \to -P -Q$ then puts the
sum-integral over $Q$ into the form shown.

Finally, we consider the $(ss)$ region for hard $P$.  Taking into account
that the hard momentum can flow through any of the three propagators, the
sum-integral can be written
\begin{equation}
\Im _{\rm sun}^{(ss)} (P) = 
3 T^2 \int_{\bf q r} {1 \over q^2 + m^2 + \sigma (q)} 
	{1 \over r^2 + m^2 + \sigma (r)} 
	{1 \over P^2 + 2 {\bf p} \!\cdot\! ({\bf q} + {\bf r}) 
		+ ({\bf q} + {\bf r})^2 + \Pi(S)} \,,
\end{equation}
where $S = -P - (0, {\bf q} + {\bf r})$.  
The $m/T$ expansion is obtained by expanding the sum-integral around 
$1 / [ (q^2 + m^2) (r^2 + m^2) P^2]$.  
The leading term is of order $m^2$ and all higher terms are of order
$g^4$ or higher.  Thus the $(ss)$ term in the sum-integral can be
written
\begin{equation}
\Im _{\rm sun}^{(ss)}(P) = 
3 T^2 I_1^2 {1 \over P^2} + {\cal O} (g^4).
\label{ISunh:ss}
\end{equation}
The complete expression for the sunset sum-integral for hard momentum
$P$, including all terms up to errors of order $g^4$, is the sum of
(\ref {ISunh:hh}), (\ref {ISunh:hs}), and (\ref {ISunh:ss}).

\subsubsection{Sunset sum-integral for soft $P$}

We next carry out the $m/T$ expansion for the sunset sum-integral with
soft external momentum $P = (0, {\bf p})$, where ${\bf p}$ is of order
$m$. 
Again, there are three momentum regions to consider:  
$(hh)$, $(hs)$, and $(ss)$.

We first consider the $(hh)$ region.  The sum-integral can be written
\begin{equation}
\Im _{\rm sun}^{(hh)}(0, {\bf p}) = 
\sumint_{QR} {1 \over Q^2 + \Pi(Q)} {1 \over R^2 + \Pi (R)} 
{1 \over (Q+R)^2 + 2({\bf q} + {\bf r}) \cdot {\bf p} + p^2 + \Pi(S)} \,,
\end{equation}
where $S = - (Q + R) - (0, {\bf p})$.  The $m/T$ expansion is obtained by
expanding the sum-integrand around $1 / [ Q^2 R^2 (Q + R)^2 ]$.  The
first few terms are
\begin{eqnarray}
\Im _{\rm sun}^{(hh)}(0, {\bf p}) &=& 
\sumint_{QR} {1 \over Q^2 R^2 (Q + R)^2} 
- (p^2 + 3m^2) \sumint_{QR} {1 \over (Q^2)^2 R^2 (Q+R)^2}
\nonumber 
\\
&& + {4 \over d} p^2 \sumint_{QR} { {\bf q}^2 \over (Q^2)^3 R^2 (Q+R)^2}
+ {\cal O} (g^4) \,.
\label{ISuns:hh}
\end{eqnarray}
The first sum-integral on the right side vanishes.

We next consider the $(hs)$ region.  There must be one propagator with
soft
momentum.  Taking into account that it can be any one of the three
propagators, the sum-integral can be written
\begin{equation}
\Im _{\rm sun}^{(hs)}(0, {\bf p}) = 
3 T \int_{\bf r} \sumint_Q 
{1 \over Q^2 + \Pi (Q)} {1 \over r^2 + m^2 + \sigma (r)} 
{1 \over Q^2 + 2 {\bf q} \cdot ({\bf p} + {\bf r}) 
	+ ({\bf p} + {\bf r})^2 + \Pi (S)} \,, 
\end{equation}
where $S = -Q - (0, {\bf p} + {\bf r})$.  The $m/T$ expansion is obtained
by expanding the sum-integrand around $1 / [(Q^2)^2 (r^2 + m^2)]$.  The
first few terms are
\begin{eqnarray}
\Im _{\rm sun}^{(hs)}(0, {\bf p}) &=& 
3T I_1 \sumint_Q {1 \over (Q^2)^2} 
- 3T \left [ I_0 + (p^2 + m^2) I_1 \right ] \sumint_Q {1 \over (Q^2)^3} 
\nonumber 
\\
&& - 3 (g_0 \mu^{- \epsilon})^4 T 
	\int_{\bf r} {\sigma_{4,-2} (r) \over (r^2 + m^2)^2} 
	\sumint_Q {1 \over (Q^2)^2} 
\nonumber 
\\
&& + {12 \over d} T \left [ I_0 + (p^2 - m^2 ) I_1 \right ] 
	\sumint_Q{{\bf q}^2 \over (Q^2)^4} + {\cal O} (g^4) \,.
\label{ISuns:hs}
\end{eqnarray}
The integrals $I_0$ vanish in dimensional regularization.

Finally, we consider the $(ss)$ region. The sum-integral in this region
can be written
\begin{equation}
\Im _{\rm sun}^{(ss)} = 
T^2 \int_{\bf q r} {1 \over q^2 + m^2 + \sigma(q)} 
	{1 \over r^2 + m^2 + \sigma (r)} 
	{1 \over s^2 + m^2 + \sigma (s)} \,,
\end{equation}
where ${\bf s} = -({\bf p} + {\bf q} + {\bf r})$.  The $m/T$ expansion is
obtained by expanding in powers of $\sigma (q)$, $\sigma (r)$, and
$\sigma (s)$.  The first few terms in the expansion are
\begin{equation}
\Im _{\rm sun}^{(ss)}(0, {\bf p}) = 
T^2 I_{\rm sun} (p) 
- 3 (g_0 \mu^{- \epsilon})^4 T^2 
	\int_{\bf q} {\sigma_{4,-2} (q) \over (q^2 + m^2)^2}
                      I_{\rm bub}(|{\bf p}+{\bf q}|)
+ {\cal O} (g^4) \,,
\label{ISuns:ss}
\end{equation}
where $I_{\rm sun} (p)$ and $I_{\rm bub}(p)$ are the sunset and bubble 
integrals defined in (\ref{Isun}) and (\ref{Ibub}).

The complete expression for the sunset sum-integral at soft momentum $P =
(0, {\bf p})$, including all terms up to errors of order $g^4$, is
the sum of (\ref{ISuns:hh}), (\ref{ISuns:hs}), and (\ref{ISuns:ss}). 
Evaluated at $p = im$, it reduces to
\begin{eqnarray}
\Im _{\rm sun}(0, {\bf p})\bigg|_{p=im} &=&
T^2 I_{\rm sun} (im) 
+ 3 T I_1 \sumint_Q {1 \over (Q^2)^2} 
- {24 \over d} I_1 m^2 T \sumint_Q {{\bf q}^2 \over (Q^2)^4} 
\nonumber 
\\
&& -2 m^2 \sumint_{QR} {Q^2+(2/d) {\bf q}^2 \over (Q^2)^3 R^2 (Q+R)^2}
\nonumber 
\\
&& - 3 (g_0 \mu^{- \epsilon})^4 \left[ 
     T^2\int_{ \bf q} {\sigma_{4,-2}(q) \over (q^2 + m^2)^2} 
	I_{\rm bub}(|{\bf p}+{\bf q}|)\bigg|_{p=im} \right. \nonumber
\\
&&\hspace{3cm}
\left.  + T\int_{\bf r} {\sigma_{4,-2} (r) \over (r^2 + m^2)^2} 
	\sumint_Q {1 \over (Q^2)^2} \right] + {\cal O} (g^4) \,.
\label{ISuns:m}
\end{eqnarray}

\subsubsection{Sum-integrals in the thermodynamic potential}

We now apply the $m/T$ expansions to the sum-integrals that appear in the
thermodynamic potential (\ref{thpot3:1}). The method should be clear
from the example of the sunset diagram, so we only give the results for
each momentum region.

For the logarithmic sum-integral in (\ref{thpot3:1}), the contributions 
from the hard and soft regions are 
\begin{eqnarray}
&& \sumint^{(h)}_P \log (P^2 + \Pi(P)) = \sumint_P \left[ \log P^2 
+ {m^2\over P^2} - {m^4 \over 2(P^2)^2} 
			+ {m^6 \over 3 (P^2)^3} \right] \nonumber
\\
&&\hspace{4cm}
+ (g_0 \mu^{- \epsilon})^4 
	\sumint_P \left[ {\Pi_{4,0}(P) + \Pi_{4,1}(P) + \Pi_{4,2}(P) 
+ \Pi_{4,3}(P) \over P^2}\right. \nonumber
\\
&&\hspace{7cm} \left.
-m^2{\Pi_{4,0}(P) + \Pi_{4,1}(P) \over (P^2)^2} \right] \nonumber
\\
&&\hspace{4cm}
+ (g_0 \mu^{- \epsilon})^8 
\sumint_P {\Pi_{8,-2}(P) + \Pi_{8,-1}(P)\over P^2} + {\cal O} (g^8) \,,
\\
&&\sumint^{(s)}_P \log(P^2 + \Pi(P)) = T(-I_0') + (g_0 \mu^{- \epsilon})^4 T 
	\int_{\bf p} {\sigma_{4,-2} (p) + \sigma_{4,0} (p) \over 
p^2 +m^2} \nonumber
\\ 
&&\hspace{4cm}
+ (g_0 \mu^{- \epsilon})^8  T 
	\int_{\bf p} \left( {\sigma_{8,-4} (p) \over p^2+m^2}
-  {\sigma^2_{4,-2}(p) \over 2(p^2 + m^2)^2} \right) + {\cal O} (g^8) \,.
\end{eqnarray}
For the other one-loop sum-integral in the thermodynamic potential, the hard 
and soft contributions are 
\begin{eqnarray}
&& \sumint^{(h)}_P {\Pi(P) \over P^2 + \Pi(P)} = \sumint_P \left[ 
{m^2\over P^2} - {m^4 \over (P^2)^2} + {m^6 \over (P^2)^3} \right] \nonumber
\\
&&\hspace{4cm}
+ (g_0 \mu^{- \epsilon})^4 
	\sumint_P \left[ {\Pi_{4,0}(P) + \Pi_{4,1}(P) + \Pi_{4,2}(P) 
+ \Pi_{4,3}(P) \over P^2}\right. \nonumber
\\
&&\hspace{7cm} \left.
-2m^2{\Pi_{4,0}(P) + \Pi_{4,1}(P) \over (P^2)^2} \right] \nonumber
\\
&&\hspace{4cm}
+ (g_0 \mu^{- \epsilon})^8 
\sumint_P {\Pi_{8,-2}(P) + \Pi_{8,-1}(P)\over P^2} + {\cal O} (g^8) \,,
\\
&&\sumint^{(s)}_P {\Pi(P) \over P^2 + \Pi(P)} = m^2 T I_1 
+ (g_0 \mu^{- \epsilon})^4 T 
	\int_{\bf p} p^2 {\sigma_{4,-2} (p) + \sigma_{4,0} (p)  
\over (p^2 + m^2)^2} \nonumber
\\ 
&&\hspace{4cm}
+ (g_0 \mu^{- \epsilon})^8  T 
	\int_{\bf p} \left( p^2 {\sigma_{8,-4} (p) \over (p^2+m^2)^2}
- p^2{\sigma^2_{4,-2}(p) \over (p^2 + m^2)^3} \right) 
+ {\cal O} (g^8) \,.
\end{eqnarray}

The two-loop sum-integral in (\ref{thpot3:1}) is the square of the tadpole 
sum-integral. It is multiplied by $g^2_0$, so to get the free energy
through order $g^7$, we need the square of the tadpole integral to order
$g^5$.  This is obtained by squaring the sum of 
(\ref{tadpole3:h}) and (\ref{tadpole3:s}), 
and keeping all terms to the desired order.

The basketball sum-integral in (\ref{thpot3:1}) involves a triple
sum-integral, so there are 4 momentum regions:  $(sss)$, $(hss)$,
$(hhs)$, and $(hhh)$.  The contribution from each of these regions is
\begin{eqnarray}
\Im _{\rm ball}^{(hhh)} &=&  
\sumint_{PQR} \left( {1 \over P^2 Q^2 R^2 (P+Q+R)^2} 
			- {4 m^2 \over (P^2)^2 Q^2 R^2 (P+Q+R)^2} \right)
+ {\cal O} (g^4),
\label{Iball-hhh}
\\
\Im _{\rm ball}^{(hhs)} &=& 
4T I_1 \sumint_{QR} {1 \over Q^2 R^2 (Q+R)^2} 
-4T\left(I_0+2m^2I_1\right)\sumint_{QR}{1\over (Q^2)^2 R^2 (Q+R)^2} \nonumber
\\
&& +{16\over d}T\left(I_0-m^2I_1\right)\sumint_{QR}{{\bf q}^2 \over (Q^2)^3 
R^2 (Q+R)^2} \nonumber 
\\
&& -4 (g_0 \mu^{-\epsilon})^4 T 
	\int_{\bf p} {\sigma_{4,-2}(p) \over (p^2 + m^2)^2} 
	\sumint_{QR} {1 \over Q^2 R^2 (Q+R)^2} 
+ {\cal O} (g^4)\,,
\label{Iball-hss}
\\
\Im _{\rm ball}^{(hss)} &=& 
6 T^2 I_1^2 \sumint_Q {1 \over (Q^2)^2} 
+ {\cal O} (g^4)\,,
\\
\Im _{\rm ball}^{(sss)} &=& 
T^3 I_{\rm ball} 
- 4 (g_0 \mu^{-\epsilon})^4 T^3 
	\int_{\bf p} {\sigma_{4,-2}(p) I_{\rm sun}(p) \over (p^2 + m^2)^2}  
+ {\cal O} (g^4)\,,
\label{Iball-sss}
\end{eqnarray}
where $I_{\rm ball}$ is the basketball integral given in (\ref{Iball}).
In (\ref{Iball-hss}), the first sum-integral and the integral $I_0$ vanish 
with dimensional regularization.

\section{Massless sum-integrals}
\setcounter{equation}{0}
\label{app:sumint}

In this Appendix, we list the massless sum-integrals that appear in the 
calculation of the 3-loop $\Phi$-derivable free energy through $7^{\rm th}$ 
order in $g$ and $m/T$. Analytic expressions are given for all the 
sum-integrals with the exception of one 3-loop sum-integral, for which the 
constant term has not been evaluated.
 
In the imaginary-time formalism for thermal field theory, a boson has
Euclidean 4-momentum $P=(p_0,{\bf p})$, with $P^2=p_0^2+{\bf p}^2$. 
The Euclidean energy $p_0$ has discrete values: $p_0=2\pi nT$, where $n$
is
an integer. Loop diagrams involve sums over $p_0$ and integrals over 
${\bf p}$. With dimensional regularization, the integral is generalized
to $d = 3-2 \epsilon$ spacial dimensions.
We define the dimensionally regularized sum-integral by
\begin{equation}
  \hbox{$\sum$}\!\!\!\!\!\!\int_P \;\equiv\; 
  \left(\frac{e^\gamma\mu^2}{4\pi}\right)^\epsilon\;
  T\sum_{p_0}\:\int {d^{3-2\epsilon}p \over (2 \pi)^{3-2\epsilon}}\,,
\label{sumint-def}
\end{equation}
where $d=3-2\epsilon$ is the dimension of space and $\mu$ is an arbitrary
momentum scale. The factor $(e^\gamma/4\pi)^\epsilon$
is introduced so that, after minimal subtraction of the poles in
$\epsilon$
due to ultraviolet divergences, $\mu$ coincides with the renormalization
scale of the $\overline{\rm MS}$ renormalization scheme.

The basic one-loop sum-integral is
\begin{eqnarray}
\sumint_P {1 \over (P^2)^n} &=&
{\zeta(2n-3+2 \epsilon) \over  8 \pi^2} 
{\Gamma(n-{3\over2} + \epsilon) \over \Gamma({1\over2}) \Gamma(n)}
\left( e^\gamma \mu^2 \right)^\epsilon
(2 \pi T)^{4-2n-2\epsilon} \,,
\label{SIn-result}
\end{eqnarray}
where $\gamma$ is Euler's constant and $\zeta(z)$ is Riemann's 
zeta function.  Most of the explicit one-loop sum-integrals 
required in our calculations 
can be derived from (\ref{SIn-result}):
\begin{eqnarray}
\sumint_P \log P^2 &=&
-{\pi^2T^4 \over 45} \left[1 + O(\epsilon) \right] \,,
\\
\sumint_P {1 \over P^2} \hspace{0.5cm} &=&
{T^2 \over 12} \left({\hat \mu\over 2}\right)^{2\epsilon}
\left[ 1 + \left( 2 + 2 \frac{\zeta'(-1)}{\zeta(-1)} \right) \epsilon 
	+ O(\epsilon^2) \right] \,,
\label{sumint:1}
\\
\sumint_P {1 \over (P^2)^2} &=&
{1 \over (4\pi)^2} \left({\hat \mu\over 2}\right)^{2\epsilon}
\left[ {1 \over \epsilon} + 2 \gamma 
	+ \left( {\pi^2 \over 4} - 4 \gamma_1 \right) \epsilon 
	+ O(\epsilon^2) \right] \,,
\\
\sumint_P {1 \over (P^2)^3} &=&
{1 \over (4\pi)^4 T^2} 
\left[ 2 \zeta(3) + O(\epsilon) \right] \,,
\end{eqnarray}
where $\hat{\mu}=\mu/(2\pi T)$. The number $\gamma_1$ is the first Stieltjes 
gamma constant defined by the equation
\begin{equation}
\zeta(1+z) = {1 \over z} + \gamma - \gamma_1 z + O(z^2).
\end{equation}
We also need one other one-loop sum-integral:
\begin{eqnarray}
\sumint_P {p_0^2 \over (P^2)^4} &=&
{1 \over (4\pi)^4 T^2} 
\left[ \zeta(3) + O(\epsilon) \right] \,.
\end{eqnarray}

The two-loop sum-integrals that are needed are 
\begin{eqnarray}
\sumint_{PQ} {1\over P^2 Q^2 (P+Q)^2} \hspace{0.5cm} &=& 0\,,
\\
\sumint_{PQ} {1 \over (P^2)^2 Q^2 (P+Q)^2} &=& 
{1 \over 2 (4 \pi)^4} \left({\hat \mu\over 2}\right)^{4\epsilon}
\left[ {1 \over \epsilon^2} 
+ \left( 1 + 4 \gamma \right) {1 \over \epsilon} \right.
\nonumber
\\
&& \hspace{3cm} \left.
+ \left( 3 + {\pi^2 \over 2} + 4 \gamma + 4 \gamma^2 
		- 8 \gamma_1 \right)
	+ O(\epsilon) \right]    \,,
\label{SI2l-2}
\\
\sumint_{PQ} {p_0^2 \over (P^2)^3 Q^2 (P+Q)^2} &=& 
{1 \over 8(4 \pi)^4} \left({\hat \mu\over 2}\right)^{4\epsilon}
\left[ {1 \over \epsilon^2} 
+ \left( {9\over2} + 4 \gamma \right) {1 \over \epsilon} \right.
\nonumber
\\
&& \hspace{3cm} \left.
+ \left({35 \over 4} + {\pi^2 \over 2} + 18 \gamma + 4 \gamma^2 
		- 8 \gamma_1 \right)
	+ O(\epsilon) \right]    \,.
\label{SI2l-3}
\end{eqnarray}
The sum-integrals (\ref{SI2l-2}) and (\ref{SI2l-3}) can be evaluated using 
the methods of Ref. \cite{Arnold-Zhai}. We will need two specific linear 
combinations of these sum-integrals:
\begin{eqnarray}
\sumint_{PQ} {P^2+(2/d){\bf p}^2 \over (P^2)^3 Q^2 (P+Q)^2} &=& 
{3 \over 4(4 \pi)^4} \left({\hat \mu\over 2}\right)^{4\epsilon}
\left[ {1 \over \epsilon^2} +\left({5\over 6}+4\gamma\right)
{1\over \epsilon} \right. \nonumber
\\
&&\hspace{2cm}
\left. +\left( {89\over 36}+{\pi^2\over 2}+{10\over 3}\gamma+4\gamma^2
-8\gamma_1 \right) +{\cal O}(\epsilon)\right]\,,
\\
\sumint_{PQ} {P^2-(4/d){\bf p}^2 \over (P^2)^3 Q^2 (P+Q)^2} &=& 
{1 \over 4(4 \pi)^4} \left({\hat \mu\over 2}\right)^{4\epsilon}
\left[ {1\over \epsilon}+\left( {19\over 6}+4\gamma\right)
+{\cal O}(\epsilon) \right]\,.
\end{eqnarray}
The three-loop sum-integrals  that are required are
\begin{eqnarray}
\sumint_{PQR}\frac{1}{P^2 Q^2 R^2 S^2}  
&=& {T^4 \over 24(4\pi)^2} \left({\hat \mu\over 2}\right)^{6 \epsilon}
\left[ {1 \over \epsilon} + \left( {91 \over 15} 
	+ 8 {\zeta'(-1) \over \zeta(-1)} - 2 {\zeta'(-3) \over \zeta(-3)}
	\right) + O(\epsilon) \right] \,,
\label{SI3l-1}
\\
\sumint_{PQR}\frac{1}{(P^2)^2 Q^2 R^2 S^2}  
&=& {T^2\over 8(4\pi)^4}
\left({\hat \mu\over 2}\right)^{6\epsilon}\left[ {1\over \epsilon^2}
+\left({89\over 6}+4\gamma+2{\zeta'(-1)\over \zeta(-1)}\right)
{1\over \epsilon}\right] \,, 
\label{SI3l-2}
\end{eqnarray}
where $S=P+Q+R$. The massless basketball sum-integral (\ref{SI3l-1}) was 
first evaluated analytically in Ref. \cite{Arnold-Zhai}.  The sum-integral 
(\ref{SI3l-2}) is more difficult. One way to evaluate it is to use the 
result for the massive basketball sum-integral in Ref.~\cite{bball}. If that 
sum-integral is expanded in powers of $m/T$ as in (\ref{Iball:exp}), the 
sum-integral (\ref{SI3l-2}) appears at order $(m/T)^2$. One can deduce the 
pole terms from the analytic expressions for the pole terms in the massive 
basketball sum-integral in Ref.~\cite{bball}. We have not evaluated the 
constant term in this sum-integral. 

The transcendental constants that appear in these sum-integrals are
\begin{eqnarray}
\gamma \ \ &=& 0.57722 \,,
\\
\gamma_1 \ &=& -0.072816 \,, 
\\
\zeta(3) &=& 1.20206 \,, 
\\
\zeta'(-1)/\zeta(-1) &=& 1.98505 \,,
\\
\zeta'(-3)/\zeta(-3) &=& 0.64543 \,.
\end{eqnarray}

\section{Integrals}
\setcounter{equation}{0}
\label{app:int}

In this Appendix, we list the integrals that appear in the calculation of 
the 3-loop $\Phi$-derivable free energy through $7^{\rm th}$ order in $g$ 
and $m/T$. The integrals that contribute through $5^{\rm th}$ order and 
those that contribute to the pole terms at $6^{\rm th}$ and $7^{\rm th}$ 
order are evaluated analytically. Some of the integrals that contribute to 
finite terms at $6^{\rm th}$ and $7^{\rm th}$ order are evaluated numerically.

In a Euclidean field theory in 3 space dimensions, 
loop diagrams involve integrals over 3-momenta. 
With dimensional regularization, the integral is generalized
to $d = 3-2 \epsilon$ spacial dimensions.
We define the measure for the dimensionally regularized integral by
\begin{equation}
  \int_{\bf p}\;\equiv\;
  \left(\frac{e^\gamma\mu^2}{4\pi}\right)^\epsilon\,
  \int {d^{3-2\epsilon}p \over (2 \pi)^{3-2\epsilon}}\,.
\label{int-def}
\end{equation}
If renormalization is accomplished by the minimal subtraction of poles
in $\epsilon$, then $\mu$ is the renormalization scale in the 
$\overline{\rm MS}$ scheme.

The basic one-loop integral for a boson with mass $m$ is
\begin{eqnarray}
I_n &=& \int_{\bf p} {1 \over (p^2+ m^2)^n} \,.
\label{In-def}
\end{eqnarray}
Differentiating with respect to $n$ and then setting $n=0$, 
we obtain the integral
\begin{eqnarray}
-I'_0 &=& \int_{\bf p} \log(p^2+ m^2) \,.
\end{eqnarray}
The analytic expression for the integral (\ref{In-def}) is
\begin{eqnarray}
I_n &=& {1  \over  8 \pi} 
{\Gamma(n-{3\over2} + \epsilon) \over \Gamma({1\over2}) \Gamma(n)}
\left( e^\gamma \mu^2 \right)^\epsilon
m^{3-2n-2 \epsilon} \,.
\label{In-result}
\end{eqnarray}
The explicit one-loop integrals required in our calculations 
can be derived from (\ref{In-result}):
\begin{eqnarray}
I_0 & = & 0 \,,
\\ 
-I_0' & = & 
-{m^3\over 6\pi} \left( {\mu \over 2 m} \right)^{2 \epsilon}
\left[1 + {8 \over 3} \epsilon 
	+ O(\epsilon^2) \right]\,,
\label{bilog}
\\ 
I_1 & = & 
-{m\over 4\pi} \left( {\mu \over 2 m} \right)^{2 \epsilon}
\left[1 + 2 \epsilon + \left( 4 + {\pi^2 \over 4} \right) \epsilon^2
	+ O(\epsilon^3) \right]\,,
\label{bi1}
\\ 
I_2 & = & 
{1\over 8\pi m} \left( {\mu \over 2 m} \right)^{2 \epsilon}
\left[1 + O(\epsilon^2)\right]\,.
\label{I2}
\end{eqnarray}

The multi-loop integrals that are needed can be conveniently expressed 
in terms of the bubble and sunset integrals defined by
\begin{eqnarray}
I_{\rm bub}(p) &=& \int_{\bf q}  {1\over q^2+m^2} {1\over ({\bf p}+
{\bf q})^2+m^2}\, ,
\label{Ibub}
\\ 
I_{\rm sun}(p) &=& \int_{\bf qr} {1\over q^2+m^2} {1\over r^2+m^2} 
	{1\over ({\bf p}+{\bf q}+{\bf r})^2+m^2} \,.
\label{Isun}
\end{eqnarray}
We need the analytic continuation of the sunset integral to $p = i m$. 
It is evaluated in Ref. \cite{Braaten-Nieto:scalar} through order 
$\epsilon^0$, but we also need the order $\epsilon$ term:
\begin{eqnarray}
I_{\rm sun}(im) & = & 
{1\over 4(4\pi)^2} \left( {\mu \over 2 m} \right)^{4 \epsilon}
\left[ {1\over\epsilon} + (6 - 8 \log 2) \right.\nonumber
\\
&&\left. + \left(24-{7\pi^2\over 6}-2\gamma+8\gamma^2-48\log 2
+8\log^2 2\right)\epsilon+O(\epsilon^2) \right]\,.
\label{Isun-im}
\end{eqnarray}
The three-loop basketball integral can be expressed as
\begin{eqnarray}
I_{\rm ball} &=& \int_{\bf p} {I_{\rm sun}(p) \over p^2+m^2} \,.
\end{eqnarray}
This integral is also evaluated in Ref. \cite{Braaten-Nieto:scalar}:
\begin{eqnarray}
I_{\rm ball}
& = & -{m\over (4\pi)^3} \left( {\mu \over 2 m} \right)^{6 \epsilon}
\left[{1\over\epsilon} + (8 - 4\log 2)
	+ O(\epsilon)\right] \,.
\label{Iball}
\end{eqnarray}
The other three-loop integral is
\begin{eqnarray}
\int_{\bf p} {I_{\rm sun}(p) - I_{\rm sun}(im) \over (p^2+m^2)^2}
& = & - {1 \over 2(4\pi)^3 m} \left[ (1 -  \log 2) + O(\epsilon) \right]
\,.
\label{dIsun2}
\end{eqnarray}
This can be obtained by differentiating (\ref{Iball}) with respect to
$m^2$
and also using (\ref{I2}) and (\ref{Isun-im}).
The integral (\ref{dIsun2}) is convergent for $d=3$, so it can also be 
calculated directly without using dimensional regularization.
In the limit $\epsilon \to 0$, the function in the numerator  is
\begin{eqnarray}
I_{\rm sun}(p) - I_{\rm sun}(im) \; \longrightarrow \;  
- {1 \over (4\pi)^2} \left[ {3 m \over p} {\rm atan} {p \over 3 m} 
		+ {1 \over 2} \log {p^2 + 9 m^2 \over 64 m^2} \right] \,.
\label{dIsun}
\end{eqnarray}

The 4-loop integral that is needed is
\begin{equation}
\int_{\bf q} {I_{\rm sun}(q) - I_{\rm sun}(im) \over (q^2+m^2)^2}
	I_{\rm bub}(|{\bf p}+{\bf q}|)\bigg|_{p=im}
={1 \over  (4\pi)^4 m^2} \left[ -0.022048 + O(\epsilon) \right].
\label{i4loop}
\end{equation}
%
This integral can be evaluated directly in $d=3$ dimensions.
The bubble integral is
\begin{eqnarray}
I_{\rm bub}(|{\bf p}+{\bf q}|)
\; \longrightarrow \; 
{1 \over  4 \pi |{\bf p} + {\bf q}|} 
	{\rm atan} {|{\bf p} + {\bf q}| \over 2 m}\,.
\end{eqnarray}
After averaging over the angles of ${\bf q}$, we can set $p = im$.
Inserting the resulting function into (\ref{i4loop}) along with
(\ref{dIsun}), we obtain a  one-dimensional integral that can
be evaluated numerically.
 
Finally there are several 5-loop integrals that are needed.
Two of them are very simple:
\begin{eqnarray}
\int_{\bf p} {[I_{\rm sun}(p) - I_{\rm sun}(im)]^2 \over (p^2+m^2)^2}
& = & {1 \over  (4\pi)^6 m} \left[ 2.52519 + O(\epsilon) \right] ,
\\
\int_{\bf p} {[I_{\rm sun}(p) - I_{\rm sun}(im)]^2 \over (p^2+m^2)^3}
& = & {1 \over  (4\pi)^6 m^3} \left[ 0.012004 + O(\epsilon) \right] .
\end{eqnarray}
They can be evaluated directly in $d=3$ dimensions
by inserting the expression (\ref{dIsun}) for the function in the
numerator.
The other 5-loop integral is 
\begin{eqnarray}
\int_{\bf p} {I_{\rm sun}(p) - I_{\rm sun}(im) \over (p^2+m^2)^2}
	{d  \ \over d m^2} I_{\rm sun}(p)
& = & {1 \over  (4\pi)^6 m^3} \left[ 0.344713 + O(\epsilon) \right] .
\label{I5loop}
\end{eqnarray}
In the limit $\epsilon \to 0$, the last factor in the integrand
reduces to
\begin{eqnarray}
{d \ \over d m^2} I_{\rm sun}(p) \; \longrightarrow \; 
- {1 \over (4\pi)^2} \left[ {3\over 2m p} {\rm atan} {p \over 3 m} \right]
\,.
\label{dIsundm}
\end{eqnarray}
The integral in (\ref{I5loop}) can be evaluated directly 
in $d=3$ dimensions
by inserting the expressions (\ref{dIsun}) and (\ref{dIsundm}) 
for the functions in the numerator.

\section{Massive Sum-integrals}
\setcounter{equation}{0}
\label{app:mass-sumint}

To calculate the $m/T$ expansion of the thermodynamic potential, we need to 
expand several massive sum-integrals in powers of $m/T$. The $m/T$ expansions 
can be obtained from those in Appendix~\ref{app:movert} simply by omitting 
terms containing the self-energy functions $\Pi_{n,k}(P)$ and 
$\sigma_{n,k}(p)$. The remaining terms in the $m/T$ expansion involve the 
sum-integrals in Appendix~\ref{app:sumint} and the integrals in 
Appendix~\ref{app:int}. 

The two massive one-loop sum-integrals that we need are 
\begin{eqnarray}
-{\cal I}_0' &=& \sumint_P\log (P^2+m^2)\,,
\\
{\cal I}_1 &=& \sumint_P{1\over P^2+m^2}\,.
\end{eqnarray}
Their $m/T$ expansions are given in (\ref{tadpole2:2}) and (\ref{logint:2}):
\begin{eqnarray}
-{\cal I}_0'  &=& T(-I_0')+\sumint_P \left[ \log P^2 +{m^2\over 
P^2} - {m^4\over 2(P^2)^2}+{m^6\over 3(P^2)^3} \right] + {\cal O}(m^8) \,,
\\
{\cal I}_1  &=& TI_1+\sumint_P\left[ {1\over P^2} - 
{m^2\over (P^2)^2} + {m^4\over (P^2)^3} \right] + {\cal O}(m^6) \,.
\end{eqnarray}
Inserting the sum-integrals from Appendix~\ref{app:sumint} and the integrals 
from Appendix~\ref{app:int}, we get
\begin{eqnarray}
-{\cal I}_0' &=& -{\pi^2\over 45}T^4\left({\hat{\mu}\over 2}
\right)^{2\epsilon}\left\{ 1-15\hat{m}^2
+60\hat{m}^3+{45\over 2}\left({1\over \epsilon}+2\gamma \right)\hat{m}^4
\right. \nonumber
\\
&&\hspace{3cm}\left.
-{15\over 2}\zeta(3)\hat{m}^6 \right\} + {\cal O}(m^8) \,,
\\
{\cal I}_1  &=& {1\over 12}T^2\left({\hat{\mu}\over 2}
\right)^{2\epsilon} \left\{ \left[1+\left(2+2{\zeta'(-1)\over \zeta(-1)}
\right)\epsilon\right]-6\left[1+(-2\log \hat{m}+2)\epsilon\right]
\hat{m} \right. \nonumber
\\
&&\hspace{3cm}\left.
-3\left({1\over \epsilon}+2\gamma \right)\hat{m}^2
+{3\over 2}\zeta(3)\hat{m}^4\right\} + {\cal O}(m^6) \,,
\label{Imass:1}
\end{eqnarray}
where $\hat{m}=m/(2\pi T)$ and $\hat{\mu}=\mu/(2\pi T)$. The errors in the 
coefficients of $\hat{m}^n$ are one order higher in $\epsilon$ than the 
last term shown explicitly.

The screening sum-integral for a particle of mass $m$ is
\begin{equation}
{\cal I}_{\rm screen} = \sumint_{QR} {1\over (Q^2+m^2)(R^2+m^2)((P+Q+R)^2
+m^2)}\bigg|_{P=(0,{\bf p}),\, p=im}\,.
\label{Iscreen}
\end{equation}
The $m/T$ expansion is obtained by omitting the terms in (\ref{ISuns:m}) 
involving $\sigma_{4,-2}(p)$:
\begin{eqnarray}
{\cal I}_{\rm screen} &=&
T^2 I_{\rm sun} (im) 
+ 3 T I_1 \sumint_Q {1 \over (Q^2)^2} \nonumber
\\
&& -2 m^2 \sumint_{QR}  {Q^2+(2/d) {\bf q}^2 \over (Q^2)^3 R^2 (Q+R)^2}
\nonumber
\\
&& - {24 \over d} I_1 m^2 T \sumint_Q {{\bf q}^2 \over (Q^2)^4}  
+ {\cal O} (m^4) \,.
\label{Iscreen:m}
\end{eqnarray}
The sum-integrals and integrals are given in Appendices~\ref{app:sumint} 
and \ref{app:int}. Keeping finite terms through order $m$ and divergent 
terms through order $m^3$, we get
\begin{eqnarray}
{\cal I}_{\rm screen} &=& {1\over 4(4\pi)^2}T^2\left({\hat{\mu}\over 2}\right)
^{4\epsilon}\left\{ \left[ {1\over \epsilon} +\left(-4\log \hat{m}
+6-8\log 2 \right) \right] \right. \nonumber 
\\
&&\hspace{3cm}
-6\left[ {1\over \epsilon}+\left(-2\log \hat{m} +2+2\gamma\right)\right]
\hat{m} \nonumber
\\
&&\hspace{3cm} \left.
- {3\over 2}\left[{1\over \epsilon^2}+\left({5\over 6}+4\gamma\right)
{1\over \epsilon}\right]\hat{m}^2 \right\}\,. 
\label{Iscreen:m2}
\end{eqnarray}

The massive basketball sum-integral is 
\begin{equation}
{\cal I}_{\rm ball} = \sumint_{PQR} {1\over (P^2+m^2)(Q^2+m^2)(R^2+m^2)
(S^2+m^2)} \,,
\label{mass-ball}
\end{equation}
where $S=P+Q+R$. The $m/T$ expansion of this sum-integral is obtained by 
dropping the terms involving $\sigma_{4,-2}(p)$ in the sum of 
(\ref{Iball-hhh})-(\ref{Iball-sss}):
\begin{eqnarray}
{\cal I}_{\rm ball} &=&  
\sumint_{PQR} \left( {1 \over P^2 Q^2 R^2 S^2} 
			- {4 m^2 \over (P^2)^2 Q^2 R^2 S^2} \right)
\nonumber
\\
&&+T^3I_{\rm ball}+6T^2I_1^2\sumint_Q {1 \over (Q^2)^2} 
\nonumber
\\
&&-8Tm^2I_1\sumint_{QR}{Q^2+(2/d) {\bf q}^2 \over (Q^2)^3 R^2 (Q+R)^2}  
+ {\cal O}(m^4) \,.
\label{Iball:exp}
\end{eqnarray}
The sum-integrals and integrals are given in Appendices~\ref{app:sumint} and 
\ref{app:int}. Keeping finite terms through order $m$ and divergent terms 
through order $m^3$, we get
\begin{eqnarray}
{\cal I}_{\rm ball} &=&  {T^4\over 24(4\pi)^2}\left({\hat{\mu}\over 2}
\right)^{6\epsilon}\left\{ \left[ {1\over \epsilon}
+{91\over 15}+8{\zeta'(-1)\over\zeta(-1)}
-2{\zeta'(-3)\over\zeta(-3)}\right]\right. \nonumber
\\
&&\hspace{3cm}
-12\left[{1\over \epsilon}-6\log \hat{m} +8-4\log 2\right]\hat{m} \nonumber
\\
&&\hspace{3cm}
-3\left[{1\over \epsilon^2}+\left({17\over 6}
+4\gamma+2{\zeta'(-1)\over \zeta(-1)}\right){1\over \epsilon}\right]
\hat{m}^2 \nonumber
\\
&&\hspace{3cm}
\left.+18\left[{1\over \epsilon^2}+\left(-2\log \hat{m} +{17\over 6}+4\gamma
\right){1\over \epsilon}\right]\hat{m}^3 \right\} \,.
\end{eqnarray} 

\end{document}